\documentclass[epj]{svjour}
\usepackage[utf8]{inputenc}
\usepackage[T1]{fontenc}
\usepackage{graphicx}
\usepackage{grffile}
\usepackage{longtable}
\usepackage{wrapfig}
\usepackage{rotating}
\usepackage[normalem]{ulem}
\usepackage{amsmath}
\usepackage{textcomp}
\usepackage{amssymb}
\usepackage{capt-of}
\usepackage{hyperref}
\usepackage{listingsutf8}
\date{}
\title{}
\hypersetup{
 pdfauthor={},
 pdftitle={},
 pdfkeywords={},
 pdfsubject={},
 pdfcreator={Emacs 26.3 (Org mode 9.1.9)}, 
 pdflang={English}}
\begin{document}

\title{Pulse-shape calculations and applications using the AGATAGeFEM software package}
\author{J. Ljungvall\inst{1}}

\institute{Université Paris-Saclay, CNRS/IN2P3, IJCLab, 91405 Orsay, France}

\abstract{A software package for modeling segmented High-Purity 
Segmented Germanium detectors, AGATAGeFEM, is presented. The choices made for geometry 
implementation and the calculations of the electric and weighting fields are 
discussed. Models used for charge-carrier velocities are described. Numerical 
integration of the charge-carrier transport equation is explained. Impact of 
noise and crosstalk on the achieved position resolution in AGATA detectors 
are investigated. The results suggest that crosstalk as 
seen in the AGATA detectors is of minor importance for the position
resolution. The sensitivity of the pulse shapes to the parameters in the 
pulse-shape calculations is determined, this as a function of position in 
the detectors. Finally, AGATAGeFEM has been used to produce pulse-shape data 
bases for pulse-shape analyses of experimental data. The results with the
new data base indicate improvement with respect to those with the standard 
AGATA data base.}

\maketitle
\section{Introduction}
\label{sec:orgd924c7c}
\label{org50e278d} 

In nuclear-structure physics, as in many other fields of research,
the development of better instrumentation and the discoveries of new
physics are closely linked. One of the most powerful techniques to
study atomic nuclei is \(\gamma\)-ray spectroscopy, often combined
with in-beam production of the species of interest. The development
and construction of new advanced radioactive-beam and stable-beam
facilities has prompted development and construction of a new
generation of \(\gamma\)-ray spectrometers. The last before present
generation of \(\gamma\)-ray spectrometers, such as EUROBALL
\cite{Simpson1997} and GAMMASPHERE \cite{Deleplanque1987}, have
limitations in terms of efficiency, resolving power, and maximum
count-rate capabilities. A way to improve the performance is to use
fully digital electronics combined with highly segmented High-Purity
Germanium detectors and perform so-called \(\gamma\)-ray tracking
\cite{Lee1999195}. Two large projects have been developed and are
presently in the construction phase. In Europe the Advanced
GAmma-ray Tracking Array (AGATA) \cite{Akkoyun201226} and in the USA
the Gamma-Ray Energy Tracking Array (GRETA) \cite{Lee20031095}. For a
more complete discussion on \(\gamma\)-ray spectrometers and their
development over time see, e.g., Eberth and Simpson
\cite{EBERTH2008283}. A recent discussion on the current performance
of AGATA can be found in Ljungvall et al. \cite{LJUNGVALL2020163297}.

The main characteristic of \(\gamma\)-ray tracking spectrometers is
the lack of Compton suppression shields. They have been replaced by
\(\gamma\)-ray tracking: The track of the \(\gamma\)-ray through the
array is reconstructed using interaction points given by pulse-shape
analysis (PSA). The tracking allows an increase of the solid angle
covered by germanium, with an increase in efficiency. The
\(\gamma\)-ray tracking also gives a high peak-to-total. For this
method to give high performance it is crucial that the PSA gives the
interaction positions with high accuracy (i.e. less than 5 mm FWHM
at 1 MeV). The accuracy of the PSA depends on how well the detector
response is known, and to a lesser degree, the algorithm used for
the PSA. Characterisation of the HPGe detectors, and associated
electronics, are hence of great importance.

Extensive work has been done to model the response of the highly
segmented germanium detectors used by AGATA and GRETA. Over the
years several software packages have been developed within the AGATA
collaboration to model the pulse shapes from segmented detectors
\cite{Schlarb20111,Schlarb2011,Matue2014,Bruyneel20161,Bruyneel2016}.
At the same time extensive efforts have been made to provide these
codes with accurate input in terms of impurity concentrations
\cite{BIRKENBACH2011176,BRUYNEEL201192}, crosstalk and electronics
response \cite{Bruyneel2009196,BRUYNEEL200999,schlarb2008simulation},
and charge-carrier mobility \cite{Bruyneel2006764}. Within the AGATA
community it is presently the ADL \cite{Bruyneel20161,Bruyneel2016}
package that is used to produce the data bases of pulse shapes
needed for pulse-shape analysis. For work related to GRETA and
GRETINA see \cite{Paschalis201344,PRASHER201750} and references
therein.

In this paper a software package written to model the segmented HPGe
germanium detectors used in AGATA is described. The main intended
use of this package is PSA development. It can however also be used
for detector characterization or pulse-shape data bases production.
A short introduction to pulse-shape calculations in semiconductor
detectors is given in section \ref{orga7bda19}. The
AGATAGeFEM package together with models and assumptions made are
described in section \ref{orgc19d29f}. An effort to characterize
where in the detector volume the parameters entering in pulse-shape
calculations have the largest influence on the pulse shapes is
presented in \ref{org39df270}. How a poorly known crystal geometry might
effect pulse-shape analysis is investigated in section \ref{orgdced297}.
The use of the software package to benchmark the effect of crosstalk
and noise on the two pulse-shape algorithms Extensive Grid Search
(EGS) and Singular Value Decomposition (SVD) is covered in section
\ref{org78194c8}. To validate the pulse shapes calculated by the
AGATAGeFEM package pulse-shape data bases used for PSA have been
calculated with it and used for PSA. Results using the AGATAGeFEM
bases are compared to results produced with bases calculated with
the ADL \cite{Bruyneel20161,Bruyneel2016}. This is presented in
section \ref{org780aa49}. Conclusions are given in section
\ref{org368ee3f}.

\section{A short introduction to signal generation in semiconductor detectors}
\label{sec:orgb3ba6dd}
\label{orga7bda19} 

The signal (referred to as "pulse shapes" in this work) generation
in all detectors based on the motion and collection of charge
carriers is calculated using the Shockley-Ramo theorem
\cite{Ramo1939,Shockley1938}. The theorem states that the induced
charge on an electrode due to moving charges is

\begin{eqnarray}
\label{eq:orga11ad58}
\frac{dQ(t)}{dt} & = & 
  e \left[ N_h \; \vec{v}_h(\vec{r}_h)\cdot\vec{W}(\vec{r}_h) - 
  N_e \; \vec{v}_e(\vec{r}_e)\cdot\vec{W}(\vec{r}_e) \right],
\end{eqnarray}

where \(\vec{W}(\vec{r}_{e,h})=-\nabla \Phi_W(\vec{r}_{e,h})\) is
the weighting field, \(N_{e,h}\) the number of charge carriers for
electrons and holes, respectively, and
\(\vec{v_{e,h}}(\vec{r_{e,h}})\) are the charge-carrier velocities.
The charge carrier velocities are functions of the electric field
\(\vec{E}(\vec{r})\). The electric field is calculated from the
electric potential as \(\vec{E}(\vec{r})=-\nabla \Phi(\vec{r})\).

The calculation of pulse shapes for a semiconductor detector begins
with solving the two partial differential equations (PDE) 

\begin{eqnarray}
\label{eq:orgebb25a1}
\nabla^2\Phi(\vec{r})&=&-\frac{\rho(\vec{r})}{\epsilon_{Ge}}
\end{eqnarray}

and

\begin{eqnarray}
\label{eq:org63f1cf5}
\nabla^2\Phi_{W}(\vec{r})&=&0
\end{eqnarray}

known as the Poisson and Laplace equations, respectively. Together
with appropriate boundary conditions they describe the electric
(Poisson) and weighting (Laplace) potentials. In equation
\ref{eq:orgebb25a1}, \(\rho(\vec{r})\) is the free charge distribution in the
detector and \(\epsilon_{Ge}\) the dielectric constant for
germanium. Boundary conditions for the electric potential are the
applied detector bias or 0 V on the conducting surfaces. For
surfaces requiring passivation the boundary condition should include
charges and possible leakage currents. As these are unknown the
approximation of natural boundary conditions is used in this work.
The weighting potential \(\Phi_{W}(\vec{r})\) is calculated by
setting the potential on the electrode of interest to 1 and to 0
on all other conducting surfaces.

One of the major difficulties when integrating equation \ref{eq:orga11ad58}
is to find the correct function for the charge-carrier velocities
\(\vec{v_{e,h}}(\vec{r_{e,h}})\). The physics and the models used in
this work is discussed in section \ref{orgb1fa605}.

To produce realistic pulse shapes for a detector the effects of the
limited bandwidth of the electronics have to be included. The effect
of the bandwidth can in principle be measured for a system using a
pulse generator. However, the approximation using an analytical
function seems sufficient and is more practical. The crosstalk
between different segments has to be modeled. Typically the
difference is made between so-called linear and differential
crosstalks. The former has mainly the capacitive coupling of the
electrodes as origin whereas the later has its origin in the
front-end electronics. The linear crosstalk can not be avoided,
whereas the differential crosstalk can be reduced using careful
engineering. The AGATAGeFEM package models both types of crosstalk.

\subsection{Charge-carrier motion in High-Purity Germanium detectors}
\label{sec:orgdcce874}
\label{orgb1fa605}

For all pulse-shape calculations the models used to describe the
charge-carrier mobilities are of crucial importance. A commonly
used function \cite{knoll2000} to describe the charge-carrier
velocity is
\begin{eqnarray}
\label{eq:orgdcb08ff}
\vec{v}(\vec{r})&=&\frac{\mu\vec{E}(\vec{r})}{\left(1+\left(E(\vec{r})/E_0\right)^\beta\right)^{1/\beta}}-\mu_n\vec{E}(\vec{r}),
\end{eqnarray} 
where \(E_0\), \(\beta\), \(\mu_{n}\), and \(\mu\) are
experimentally adjusted parameters. This parameterization is valid
when the electric field is parallel to one of the symmetry axes of
the Germanium crystal (<100>, <110>, or <111>). The charge-carrier
velocity is only parallel to the electric field when the electric
field is parallel to a symmetry axes. For germanium crystals cooled
down to liquid nitrogen temperatures (\(\approx-175^{\circ}\)
Celsius) several models have been developed for the electron
mobilities. For AGATAGeFEM the model of Nathan \cite{Nathan1963} is
used. It treats the anisotropy of the electron drift velocity
observed in germanium with high accuracy with the formalism
described in Mihailescu et al. \cite{Mihailescu2000350}.

\begin{figure}[htbp]
\centering
\includegraphics[width=0.5\textwidth]{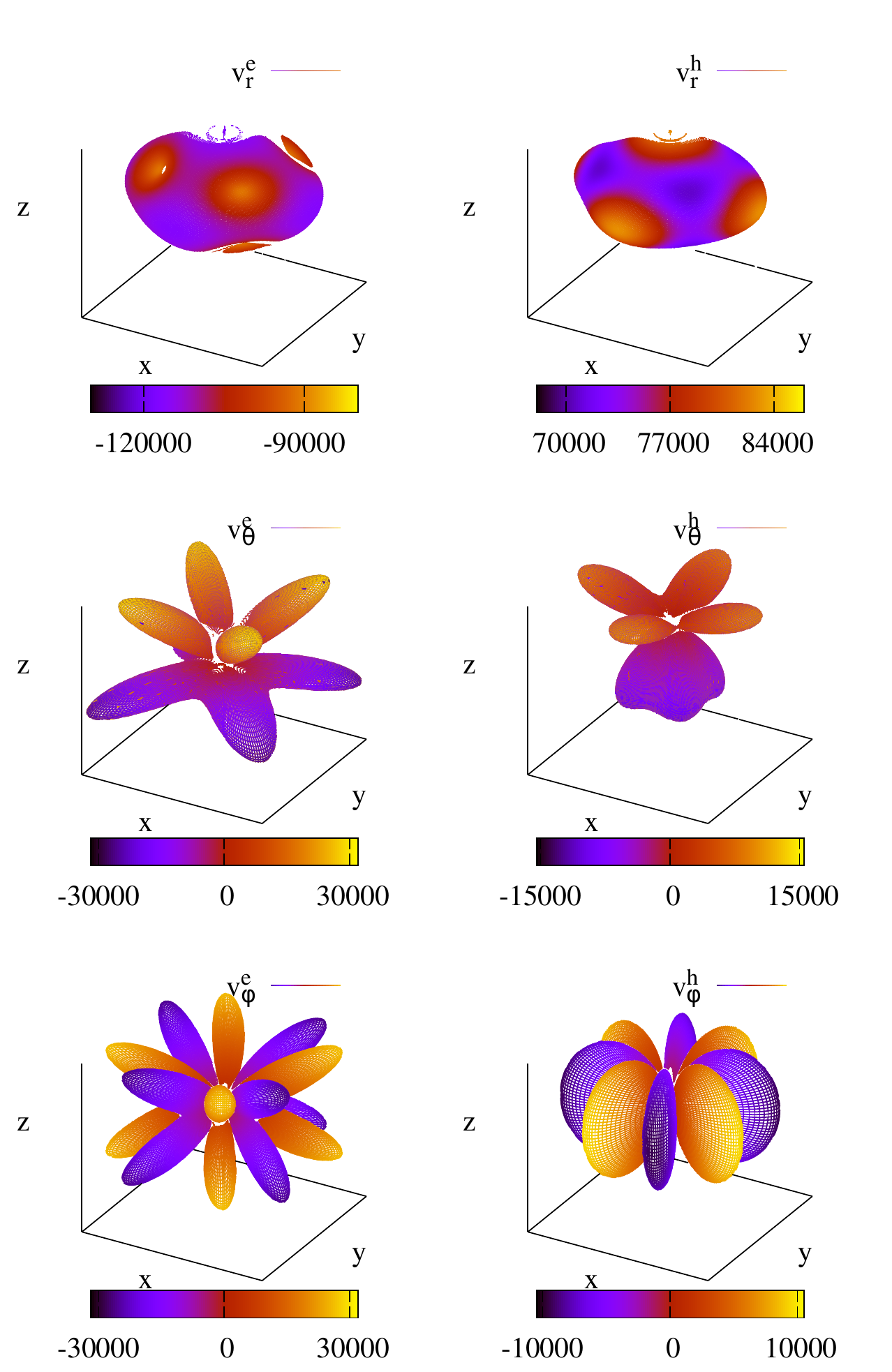}
\caption{\label{fig:org89a7f7c}
The left (right) column shows the charge-carrier velocity as a function of the direction of the electric field for electrons (holes). The three rows show the \(\hat{r},~\hat{\theta},~\hat{\phi}\) components of the velocity, respectively.}
\end{figure}

For the hole mobility B. Bruyneel et al. \cite{Bruyneel2006764} have
developed a model based on the so-called "streaming motion"
concept. The holes are accelerated to a threshold energy, they emit
an optical phonon losing most of their energy, are re-accelerated
in the applied electric field to the threshold energy, and so on.
In this work a different approach has been used. Here it is assumed
that the variation in carrier velocity as a function of the
electric field can be described by the fraction of holes populating
the light-hole band and the heavy-hole band and a field dependent
relaxation time. The anisotropy is given by the effective masses
the second derivative of the energy of the hole bands. Despite the
much higher the much higher energy of the light-hole band as
compared to the heavy-hole band the model reproduces experimental
data for hole drift velocities \cite{LjungvallThesis}. For the holes
the surfaces of equal energy in the conduction bands are not
ellipsoids, which means that the reciprocal effective mass tensor
will depend on the direction of the wavevector \(\vec{k}\). Here
the assumption is made that the wavevector is parallel to the
applied electric field. The hole energy functions read
\cite{kittel1996}
\begin{eqnarray}
\label{eq:org59f23ac}
\epsilon_h(k)&=&Ak^2\pm\left[B^2k^4+C^2\left(k_x^2k_y^2+k_y^2k_z^2+k_z^2k_x^2\right)\right]^{1/2},
\end{eqnarray}
where the positive (negative) sign is for the light (heavy)-hole band.
Using equation
\begin{eqnarray}
\label{eq:orgddbb857}
\left(\frac{1}{m^*}\right)_{\mu\nu}&=&\frac{1}{\hbar^2}\frac{d^2\epsilon(k)}{dk_{\mu}dk_{\nu}}\equiv\bar{\bar{\Gamma}}
\end{eqnarray}
to calculate the reciprocal effective mass tensor, we have
\begin{eqnarray}
\label{eq:orgb0f0d04}
\vec{v}_h&=&q\mathcal{T}\left(E\right)\left[\mathcal{F}\left(E\right)\bar{\bar{\Gamma}}^{heavy}_{h}+\left(1-\mathcal{F}\left(E\right)\right)\bar{\bar{\Gamma}}^{light}_{h}\right]\vec{E}.
\end{eqnarray}
Comparing equation \ref{eq:orgb0f0d04} with
\begin{eqnarray}
\label{eq:org6adfb30}
\vec{v}&=&qt\bar{\bar{\Gamma}}\vec{E},
\end{eqnarray}
the factor \(\mathcal{T}(E)\) in the latter equation corresponds to
\(t\) in the former and is considered an electric-field dependent
relaxation time. \(\mathcal{F}\left(E\right)\) is the fraction of
the holes moving in the heavy-hole band and is also field
dependent. As in the case of electrons, equation \ref{eq:orgdcb08ff} can now
be used to calculate the hole drift velocities in the <100> and
<111> directions. Using these velocities
\(\mathcal{T}\left(E\right)\) and \(\mathcal{F}\left(E\right)\) are
obtained at the electric-field strength in question. A big
difference for holes as compared to electrons is that the
reciprocal effective mass tensor changes with the direction of the
electric field. In figure \ref{fig:org89a7f7c} the anisotropy of
charge carrier transport is illustrated. Without anisotropy the top
row would show perfect spheres in one color and the second and
third rows would show zero velocities. The deficiency of the
model can be noted from the \(v^{h}_{\varphi}\) shown in the
bottom right corner. There should be no anisotropy in any of the
<100>, <110>, or <111> directions as these are symmetry axes in
germanium. This is different for the electrons shown in the left
column. 

Neither the drift model for electrons nor for holes takes into
account the effects of crystal temperature or impurity
concentrations on the charge-carrier drift velocities although
these effects modify the drift velocities \cite{Mei2016}. The effect
of varying the hole-drift velocity was studied within the GRETINA
\cite{PRASHER201750} and AGATA collaborations \cite{Lewandowski2019}.
In these works it is concluded that the position resolution is not
limited by the hole mobility models. When the hole mobility is
varied within reasonable limits it is difficult to separate the
effects from those coming from the front-end electronics. Numerical
values used in this work for the mobility parameters are presented
in table \ref{tab:org047b7d6}.

\section{The AGATAGeFEM software package}
\label{sec:org2a2403b}
\label{orgc19d29f}

Several software packages have been developed to calculate pulse
shapes from the High-Purity Germanium detectors used in AGATA.
Examples from the AGATA collaboration are MGS \cite{Medina,Matue2014},
JASS \cite{Schlarb2011,Schlarb2011,schlarb2008simulation} and ADL
\cite{Bruyneel20161,Bruyneel2006764,Bruyneel2016}. Although differing
in details they all have in common the use of finite difference PDEs
solvers for the electric field and the weighting potentials in the
detector. However, the complex shapes of the AGATA crystals are not
well reproduced using a finite difference scheme with rectangular
grids. This is a problem that can be circumvented using finite
element methods (FEM). Another strong point of FEM is that the
solution is an approximation of a function describing the electric
field and not the electric field at certain points. This removes the
need to interpolate between grid points as the solution is defined
on the entire volume of the detector. It is beyond the scope of
the present work to describe FEM and the reader is referred to
\cite{BrennerFEM} and references therein.

The AGATAGeFEM package written in C++, uses high-quality open-source
FEM software to calculate the electric and weighting potentials of
AGATA type germanium detectors. For the charge-carrier transport the
ordinary differential equation solvers of the Gnu Scientific Library
\cite{gsl2010} are used. The geometry is described to machine
precision for charge transport and mesh generation. Earlier versions
of the program used mainly the FEM library dealii
\cite{BangerthHartmannKanschat2007,dealII90}. This is a very flexible
code that allows an iterative refinement of the FEM mesh in a very
simple way. However, the mesh cell geometry is limited to
quadrilaterals and hexahedra. From the point of view of solving the
partial differential equations this is a good choice. However, in
AGATAGeFEM the solutions of the Poisson and Laplace equations are
not projected down to a regular grid when used in the charge
carrier-transport process. This is also the case for calculations of
the induced signals via the Shockley-Ramo theorem. The idea Behind
is that the mesh refinement procedure tells where a high granularity
is needed and all projection to a regular grid deteriorates this
information. The problem is then to find the correct cell in an
irregular mesh. The curved boundaries of hexahedra cells make these
calculations complicated. As a result the first version of
AGATAGeFEM was capable of calculating about 2-3 pulse shapes/s.
Sufficient to calculate a basis for PSA it is far from enough for
using AGATAGeFEM in the fitting of parameters in the pulse-shape
calculations or to use it in a complete Monte Carlo simulation
chain. The FEM part of the program was therefore changed to the
libmesh library \cite{Kirk2006237}. It uses tetrahedra with each side
defined by three points, speeding up finding the correct mesh cell.
The AGATAGeFEM is further more restrained to the use of only linear
bases functions in the solution. This way the calculations are a
factor of almost 100 faster while reproducing while reproducing the
results using dealii.

AGATAGeFEM is fully parallelized with threads and the \href{https://www.mcs.anl.gov/research/projects/mpi/}{MPI} interface.
This applies to the field calculations and the pulse-shape
calculations. Parallelism is also used while fitting the pulse-shape
calculations parameters. To minimize the \(\chi^2\) Minuit and
Minuit2 \cite{Hatlo2005} are employed. AGATAGeFEM further has
interfaces allowing calculating and displaying fields and pulses
from the ROOT \cite{ANTCHEVA20092499} interpreter interface. This
inside the chosen detector geometry if wanted. It has further a very
simple server client mechanism allowing other programs to ask the
server to calculate pulse shapes for it.

The AGATAGeFEM package also contains miscellaneous codes for 
\begin{itemize}
\item applying pre-amplifier response
\item crosstalk
\item re-sample pulse shapes
\item compare pulse shapes
\item calculate pulse shapes from the output of the AGATA geant4 MC
\cite{Farnea2010331}
\item create data bases for PSA.
\end{itemize}

\subsection{AGATA Detector model}
\label{sec:orgc122a28}
\label{org8f11513}
\subsubsection{Geometry}
\label{sec:orge5959ad}
\label{org3ba1838}

\begin{figure}[htb]
\begin{center}
\includegraphics[width=.9\linewidth]{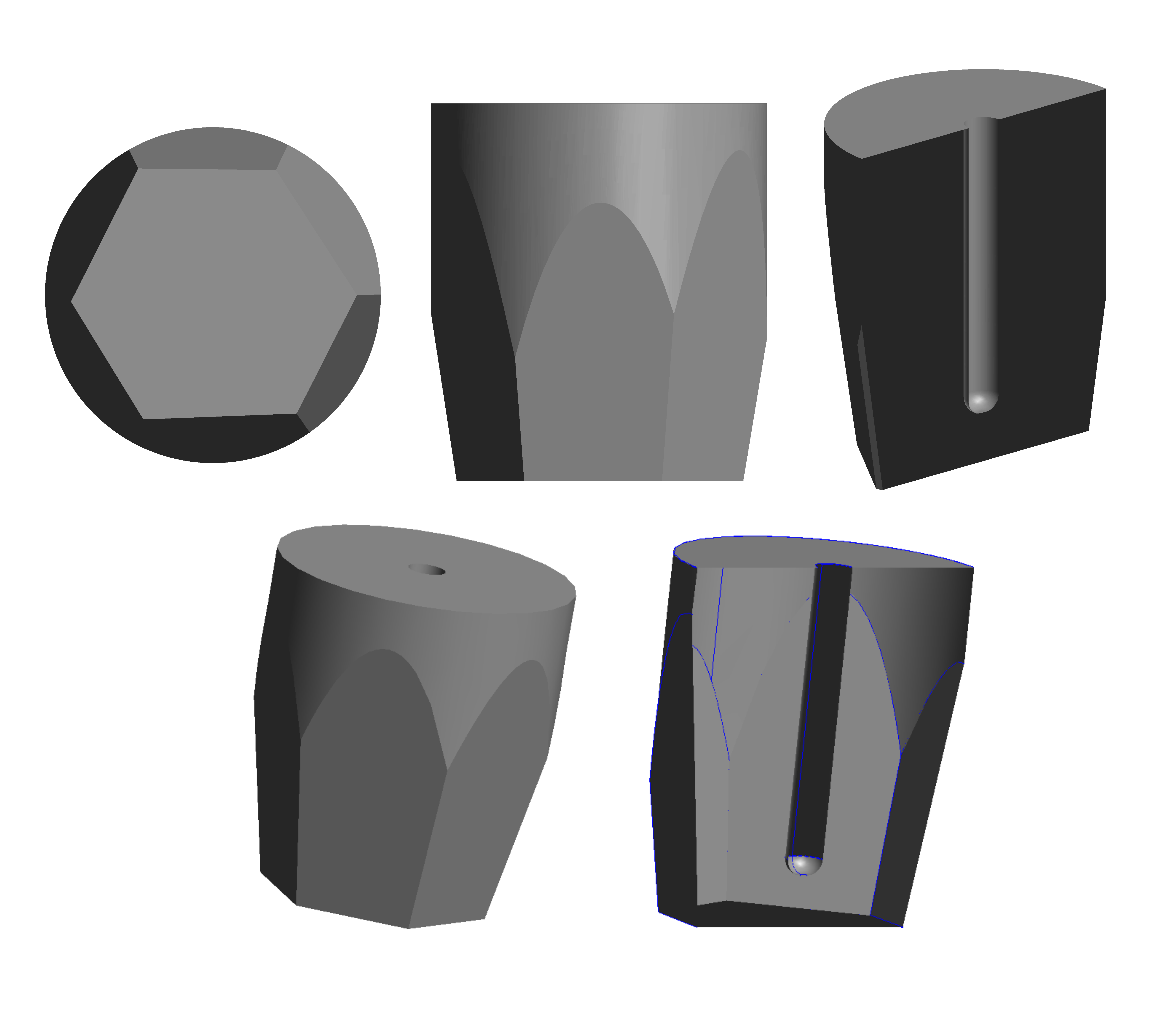}
\end{center}
\caption{\label{orgd11b85f}
Top row, three views of an "A" type AGATA crystal. Bottom left shows a symmetric crystal. Bottom right, a "C" type AGATA crystal with half the volume hidden to show the central contact.}
\end{figure}

The AGATA crystals are \(90\) mm long and have an diameter of \(80\)
mm. They were produced in four different shapes. A symmetric
hexagonal shape for three prototypes and three different
non-symmetric hexagonal shapes for use in the AGATA
\cite{Akkoyun201226}. For the generation of the FEM mesh OpenCASCADE
models of the detectors are generated. For the charge transportation
the detector geometries are implemented in C++ as the union of a
cylinder and six planes or using the CSG geometry of geant4
\cite{Agostinelli2003250}. The hole corresponding to the central
contact has several parameters allowing it shape and orientation to
be varied. These are the radius of the hole, the radius at the
bottom of the hole joining the side of the bore hole with the bottom
of it, and translation and rotation of the axes of the hole. The two
different geometrical models of the detectors are equivalent.
Examples of the geometries are shown in figure \ref{orgd11b85f}.

\subsubsection{Calculations of the electric fields and weighting fields}
\label{sec:org23af97c}
\label{orgd39e55d} AGATAGeFEM uses a total of 40 fields for
calculating the pulse shapes. Thirty seven of these are the
weighting fields for the 36 segments and the central contact.
Except for the central contact which is trivial, these are defined
either using the limiting depth values and start and stop angles
of a segment or using the intersection between the detector
surface and four planes. The segments do not have to cover the
entire surface of the detector. This is intended for modeling the
gaps between the segment contacts on the outer surface
\cite{Aydin2007}. Presently no implementation of suitable boundary
conditions for the electric field exists in AGATAGeFEM
limiting the value of this option.

When solving the charge transport equations, AGATAGeFEM uses three
fields to calculate the electric field. The first one is the
solution to the Poisson equation with \(0\) V on the surface of
the detector and \(V_{bias}\) V on the central contact, and the
nominal impurity concentration. The choice to include charge
impurities here was made to maximize the benefits of mesh
refinement. Representative values for the charge impurities are
presented in table \ref{tab:orge809380} in appendix \ref{org56d91ba}. The second
field is the solution of the Poisson equation assuming \(0\) V on
both the surface and the central contact but with an impurity
contribution of \(1\) at the front of the detector that decreases
linearly as a function of depth to the back of the detector where
it is \(0\). The third and final field is like the second one but
reversing the slope of the impurity concentration. The use of
three fields allows varying the effective impurity concentration
and its effect on the electric field in the detector without
recalculating the electric field. Solving the Poisson equation when
varying the impurity concentration is too computationally intensive
to allow the fitting of detector parameters to experimental
signals.

For solving the Laplace equation and the Poisson equation
AGATAGeFEM uses the libmesh library \cite{Kirk2006237}, with the
meshes generated with gmsh \cite{Geuzaine09}. As a first step the
Laplace or the Poisson equation is solved using a uniform mesh
with an average cell size of 2 mm. This solution is then used to
estimate the largest acceptable geometrical approximation of the
mesh in order to ensure an error on the field of less than one per
mil of the maximum value of the field. This step is then followed
by repeated steps of refinement of the mesh based on an estimate
of the local error of the solution \cite{Kelly1983} until the field
is described by at least \(2.5*10^{5}\) degrees of freedom. A
limit that gives a good approximation of the fields (see section
\ref{org780aa49}). The final step for each of the 40 fields is to
create look-up tables on a 2x2x2 mm\(^3\) grid over the detector
volume to allow fast access to the correct mesh cell when
evaluating the fields.

\subsubsection{Solving the charge transport equation}
\label{sec:orgc6f30ca}
\label{org34cf3e2} The detector pulses are calculated by
first transporting the point representing electrons and the point
representing holes from the point of the \(\gamma\)-ray
interaction until they reach the boundary of the detector volume
using
\begin{eqnarray}
\label{eq:org0d2574b}
  \frac{d\vec{r_{e,h}}}{dt}&=&\vec{v_{e,h}}\left(\vec{E}\right).
\end{eqnarray}
The equations are solved separately for the holes and the
electrons using an solver algorithm with an adaptive time step.
AGATAGeFEM allows the user to choose between any of the possible
algorithms provided by GSL, but the default choice it the embedded
Runge-Kutta Prince-Dormand method \cite{DORMAND198019}. The paths of
the charge carriers are calculated with an adaptive time step and
sampled at a chosen frequency, by default 100 MHz. As the charge
carriers approach the boundary of the detector the sampling
frequency is adapted to allow an accurate description of the pulse
shape as a \(10\) ns time step typically gives a path that ends
outside the detector.

In the next step the charge on electrode \(i\) is calculated using
\begin{eqnarray}
  Q_i\left(t\right)&=&q\left(\Phi^i_W(\vec{r_e}(t))-\Phi^i_W(\vec{r_h}(t))\right) 
\end{eqnarray}
for all \(37\) signals. 

\subsubsection{Convolution of the signals with the transfer function of the electronics}
\label{sec:orgb440e64}
\label{org9556889} 
\begin{figure}[htb]
\centering
\includegraphics[width=0.5\textwidth]{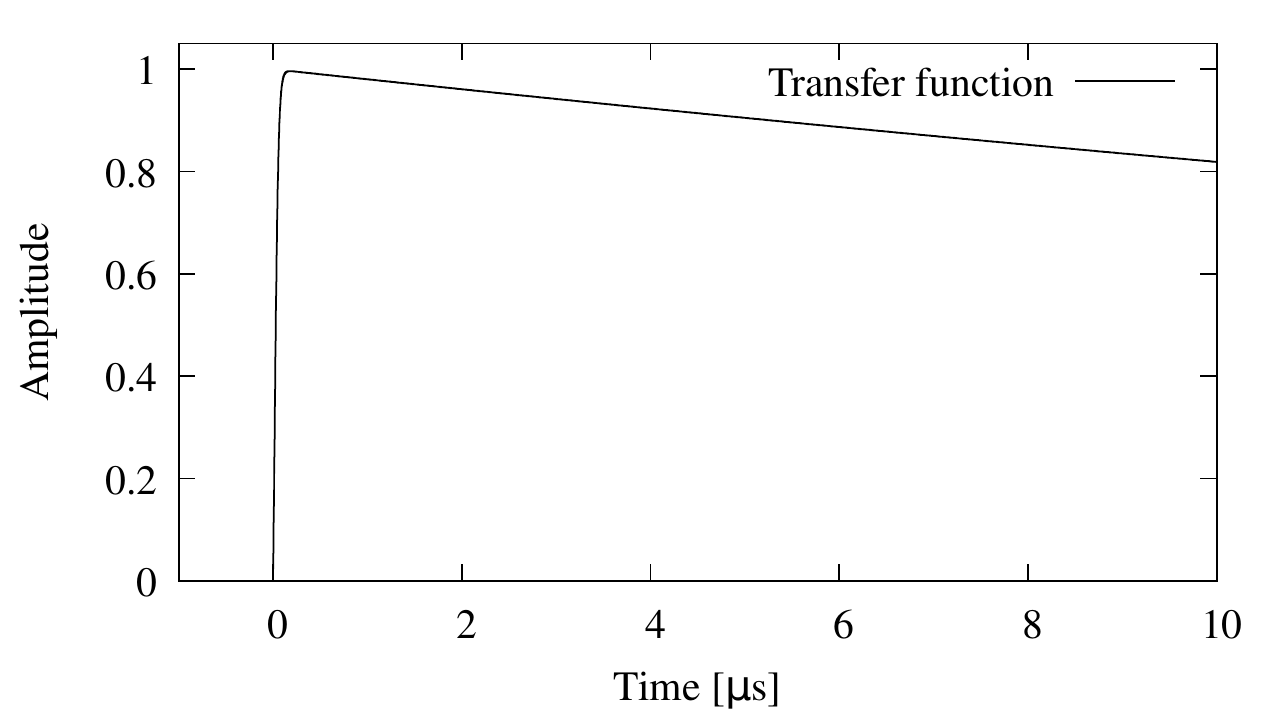}
\caption{\label{fig:org51b0f8f}
Transfer function of the electronics shown in time domain.}
\end{figure}

The signals can be
convoluted with the response of the electronics. This is not done
when generating a basis for PSA. In this work the response of the
electronics have been modeled by the function
\cite{Schlarb20111,Schlarb2011} shown in figure \ref{fig:org51b0f8f}. A
convolution (in time-domain) is made for all 37 calculated
signals. A possible improvement is to use a function with a
slower rise time for the central contact. However, the convolution
is mainly used for PSA development where AGATAGeFEM is used both
for generating the bases and the signals that are analyzed so the
exact form of the function is of minor importance.

\begin{figure}[htb]
\centering
\includegraphics[width=0.5\textwidth]{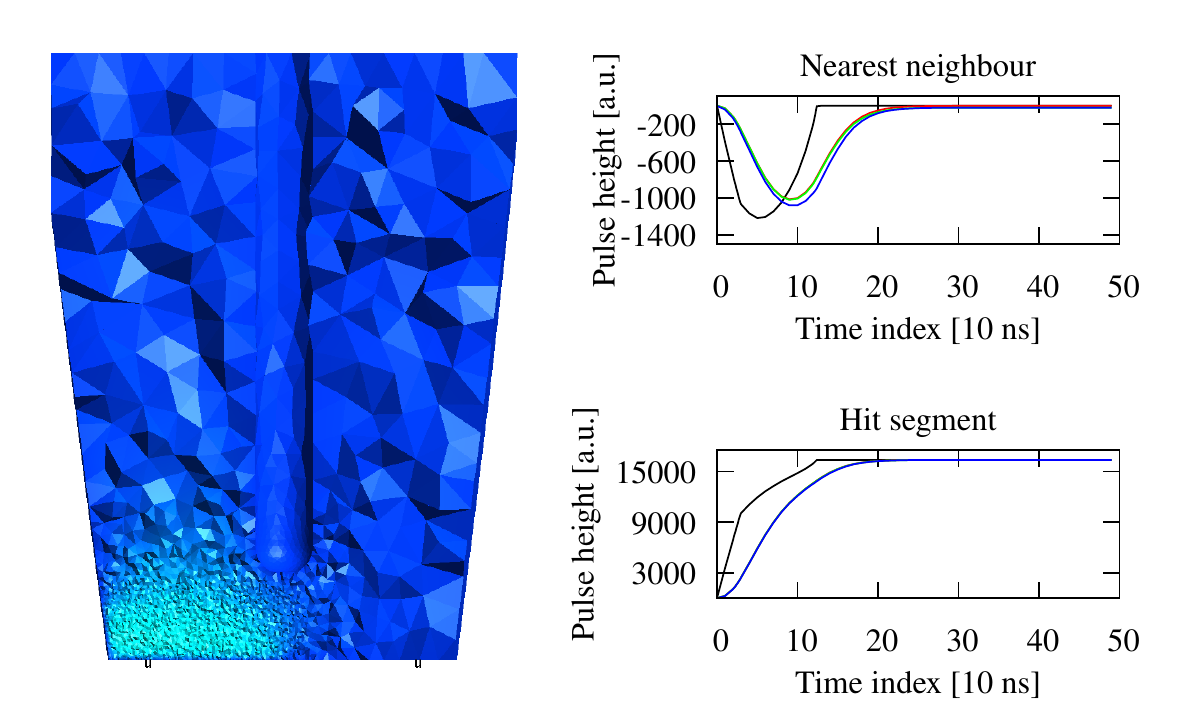}
\caption{\label{fig:org1617d21}
Left: Example of a front segment weighting field. Shown is also the mesh used for solving the Poisson equation. Right: Examples of net-charge signals and transient signals with and without convolution with the transfer function. The modulating effect of the response of the electronics is clearly seen (red shape). The effect of linear (green shape) and derivative (blue shape) crosstalk is also shown.}
\end{figure}

The effect of both linear and differential crosstalk can be
included in the transfer function. For in-depth discussion
concerning crosstalk in segmented germanium detectors, see
\cite{Bruyneel2009196,Wiens2010223}. An example of the signals
calculated with and without response function is given in figure
\ref{fig:org1617d21}. The effects of linear and derivative crosstalks
are also shown.

\section{Investigation of the sensitivity of signal shapes to detector parameters}
\label{sec:orgb9a3ff2}
\label{org39df270} 
\begin{figure}[htb]
\centering
\includegraphics[angle=0,width=0.5\textwidth]{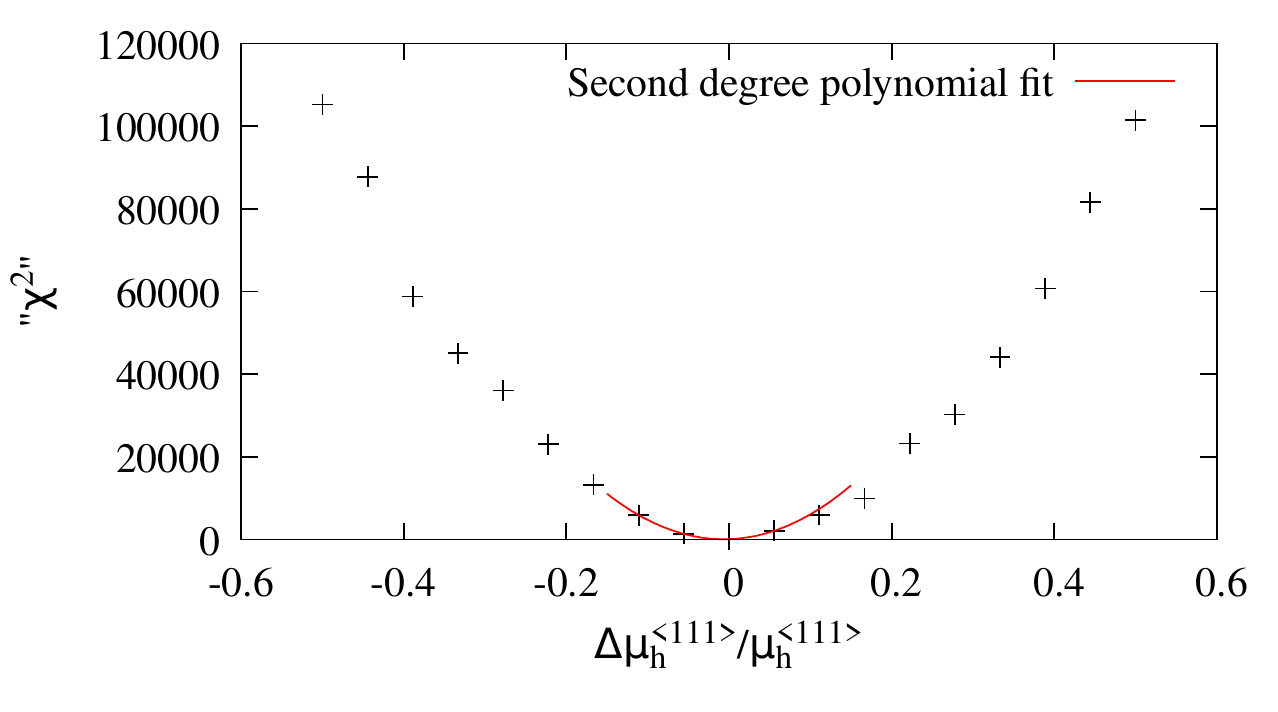}
\caption{\label{fig:org06831bb}
Example of how the sensitivity of the pulse shapes to a parameter (in this case \(\mu_{h}^{<111>}\)) is extracted.}
\end{figure}

To understand the influence of the different
detector parameters on the shape of the signals the sensitivity of
the pulse shapes to each parameter was calculated for a large number
of points inside a detector of symmetric type. As the absolute value
of the different parameters vary over many orders of magnitude 
normalized dimensionless parameters was used to estimate this
sensitivity. The sensitivity was evaluated as the second derivative
of a \(\chi^2\) at zero fractional variation of the parameter in
question. It is extracted by fitting a second degree polynomial and
is hence the curvature of the \(\chi^2\) function. The
\(\chi^2\) is calculated using the original pulse shape and a pulse
shape calculated using the changed parameter. In figure
\ref{fig:org06831bb} the extraction of the sensitivity is shown for
the hole mobility in the <111> direction.

\begin{figure}[htb]
\centering
\includegraphics[angle=0,width=0.5\textwidth]{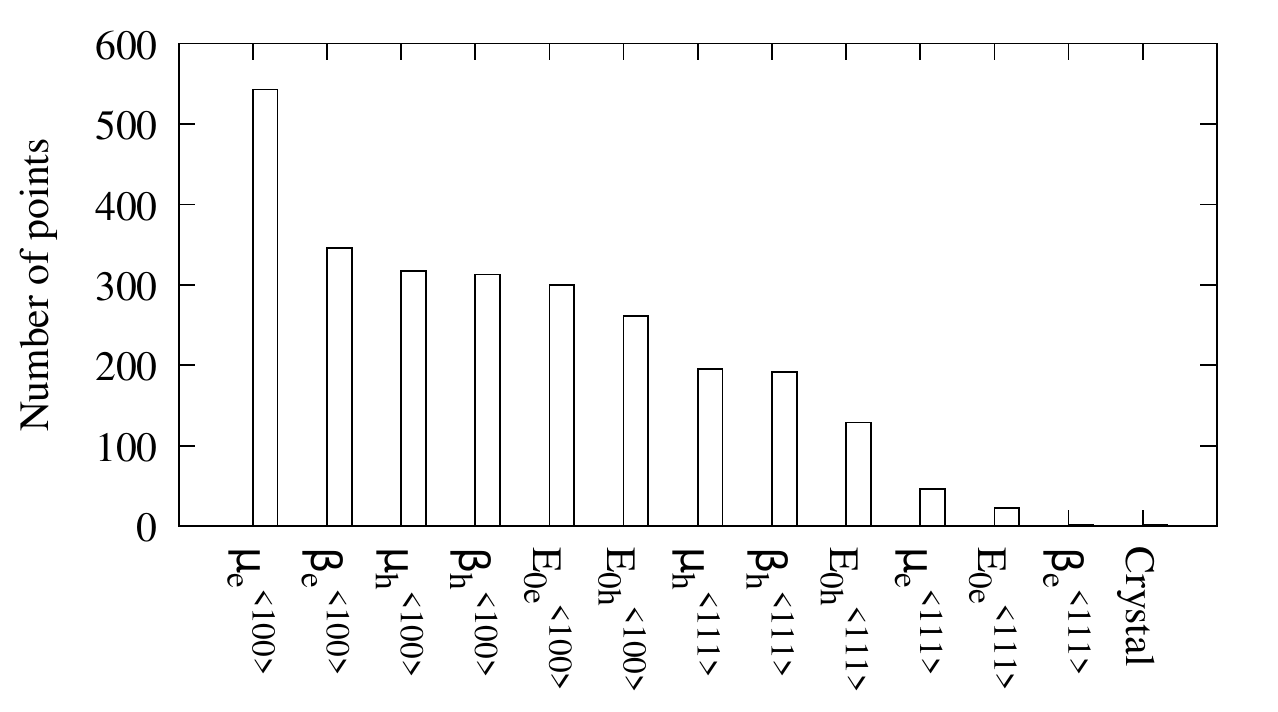}
\caption{\label{fig:org17591c2}
Number of points in the crystal where a parameter has the largest influence on the pulse shapes.}
\end{figure}

In figure \ref{fig:org17591c2} is shown for how many positions in the
detector the pulse shapes are most sensitive to each parameter. It
can be seen that at most points it is the parameters that control
the velocity of the charge carriers in the <100> direction that are
dominating. It is not surprising as in the coaxial part of the
detector the charge transport is never in a <111> direction. Worth
noting is that the hole mobility parameters for the <111> direction
have a large impact at more positions than the <111> direction
parameters for the electrons. This can also be understood
geometrically as the paths close to the <111> direction with an
overlapping large weighting potential are dominated by hole
transport occurring at the corners of the front face of the
detector. Crystal orientation only dominates at one position, a
consequence of the definition and its evaluation at zero change.

\begin{figure}[htb]
\centering
\includegraphics[angle=0,width=0.5\textwidth]{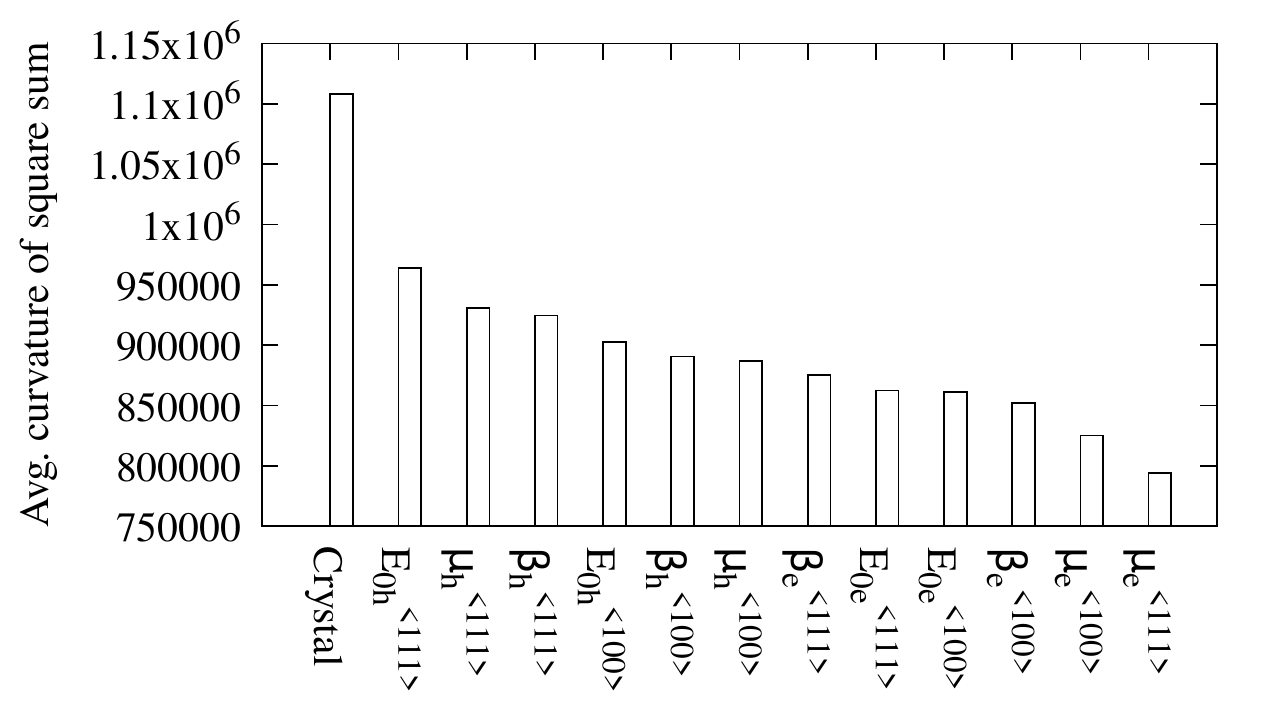}
\caption{\label{fig:orgee41550}
The sensitivity of pulse shapes to the parameters used in pulse-shape calculations averaged over the points where the pulse shapes are most sensitive to the respective parameter.}
\end{figure}

Figure \ref{fig:orgee41550} shows the average sensitivity of the pulse shapes
to a parameter at the positions dominated by the parameter. One can
notice that the highest average sensitivity is found for the crystal
orientation followed by the parameterization of the hole mobility in
the <111> direction. It can be understood as these parameters are
the most likely to change in which segment the charges are
collected.

\begin{figure}[htb]
\centering
\includegraphics[angle=0,width=0.5\textwidth]{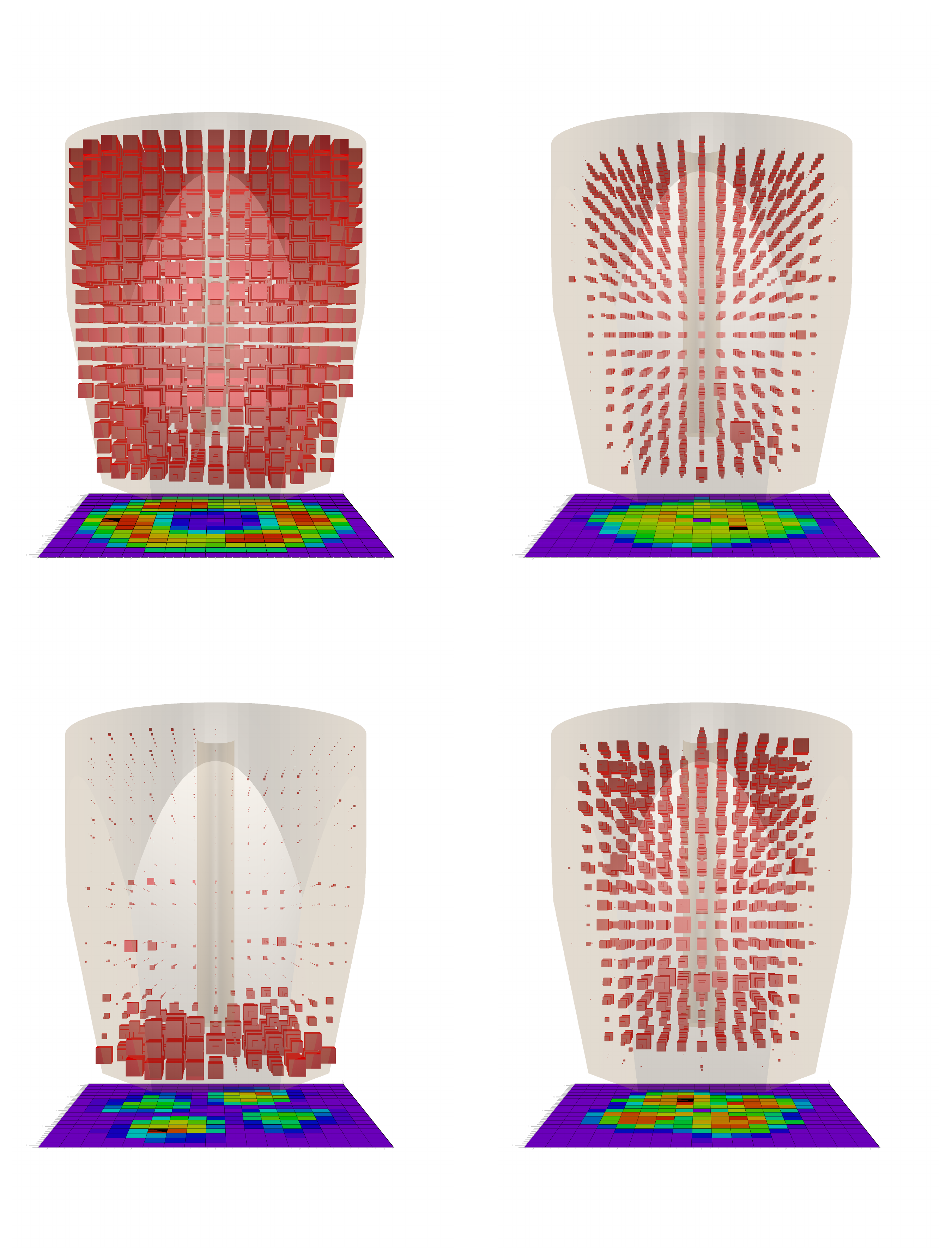}
\caption{\label{fig:org2159fc8}
The sensitivity of the pulse shapes to the \(\mu_{e,h}^{<100>,<111>}\) parameters as a function of position.  The size of the cubes are proportional to the sensitivity. Note that the cube sizes are not comparable between the figures. Top left:  \(\mu_{e}^{<100>}\) Top right: \(\mu_{h}^{<100>}\) Bottom row left: \(\mu_{e}^{<111>}\) Bottom right: \(\mu_{h}^{<111>}\)}
\end{figure}

In figure \ref{fig:org2159fc8} the relative sensitivity as a
function of position in the detector volume is shown for the
\(\mu_e^{<100>}\) parameter. It is homogeneous inside the volume
although the projection on the XY plane shows that, apart from the
volume effect, there is an increase in sensitivity close to the
<100> directions. This is expected as the charge carrier velocity
mainly depends on parameters for that crystal axes at these
positions.

A similar pattern can be seen for the \(\mu_h^{<100>}\) parameter in
figure \ref{fig:org2159fc8}, but with the maximum shifted towards
lower radii corresponding to pulses in which the hole drift
contributes more to the pulse shapes. The situation is different for
the parameters \(\mu_e^{<111>}\) and \(\mu_h^{<111>}\), also shown
in figure \ref{fig:org2159fc8}. For the electrons the pattern is
easily understood, i.e. parameters concerning the <111> direction
show sensitivity in the region where charge transport is parallel to
the <111> direction.

For the holes the pattern is not reflecting the <111> direction in
the crystal. This is not imposed by the model, unlike for the model
of the electrons. When the electric field is parallel to a symmetry
axis of the crystal the charge carriers move, by symmetry arguments,
parallel to the field and the axis. This can also be seen in figure
\ref{fig:org89a7f7c} where the \(\varphi\) component of the hole
velocity is non zero in the xy-plane <100> directions. Imposing this
symmetry on the hole velocity model is planned for future work, but
as shown in section \ref{org780aa49} this deficiency does not seem to
degrade the results.

\section{Impact of imperfectly known crystal geometry on PSA}
\label{sec:org72f8676}
\label{orgdced297} As the exact geometry (here considering contact
thickness dead layers etc as a part of the geometry) is imperfectly
known it is interesting to investigate its possible impact on the
pulse shapes. A different geometry was used to generate the basis
for PSA. The influence of an imperfect geometry has been
investigated for three different cases, ranging from an extreme
(unrealistic) case to a small error on the used front-face
segmentation.

\begin{figure}[htbp]
\centering
\includegraphics[width=.9\linewidth]{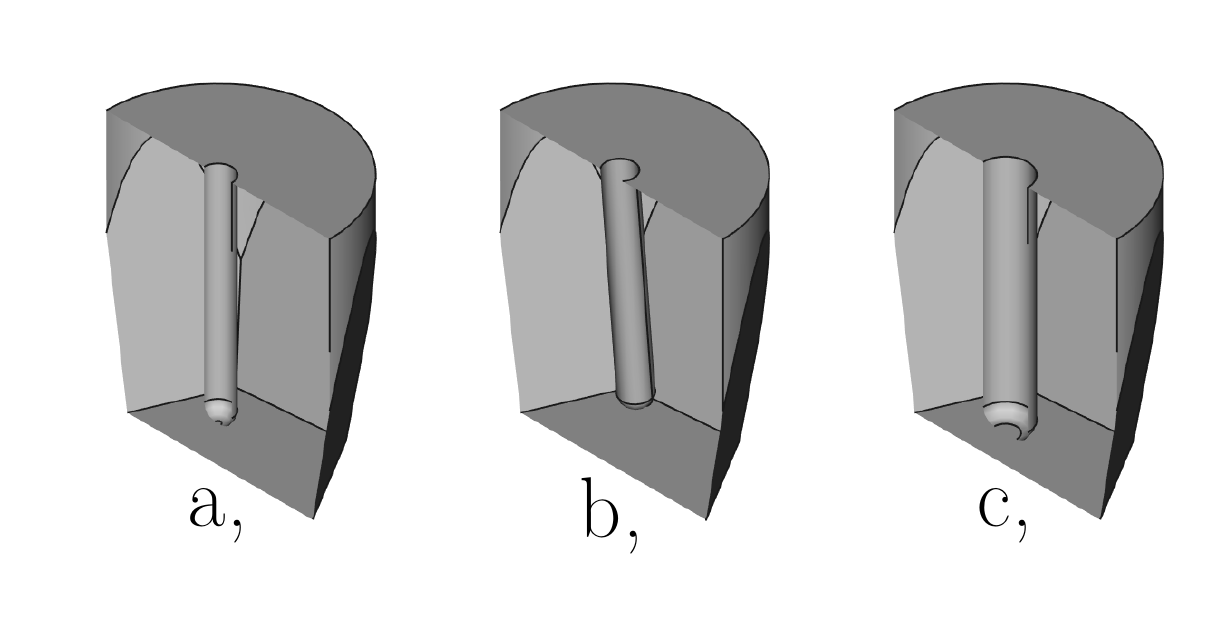}
\caption{\label{fig:orgf582c04}
The three different geometries used for investigating the impact of an imperfect geometry on the pulse-shape data bases. In picture a, the nominal geometry of a capsule type A is shown. In b the bore hole for the central contact has been displace 5 mm and with an angle of \(\phi=.2\) radians  and \(\theta=.1\) radians, respectively. In figure c the bore hole has been enlarged with 3 mm, this as an (extreme) example of the lithium drifted central contact.}
\end{figure}

\begin{figure}[htbp]
\centering
\includegraphics[width=.9\linewidth]{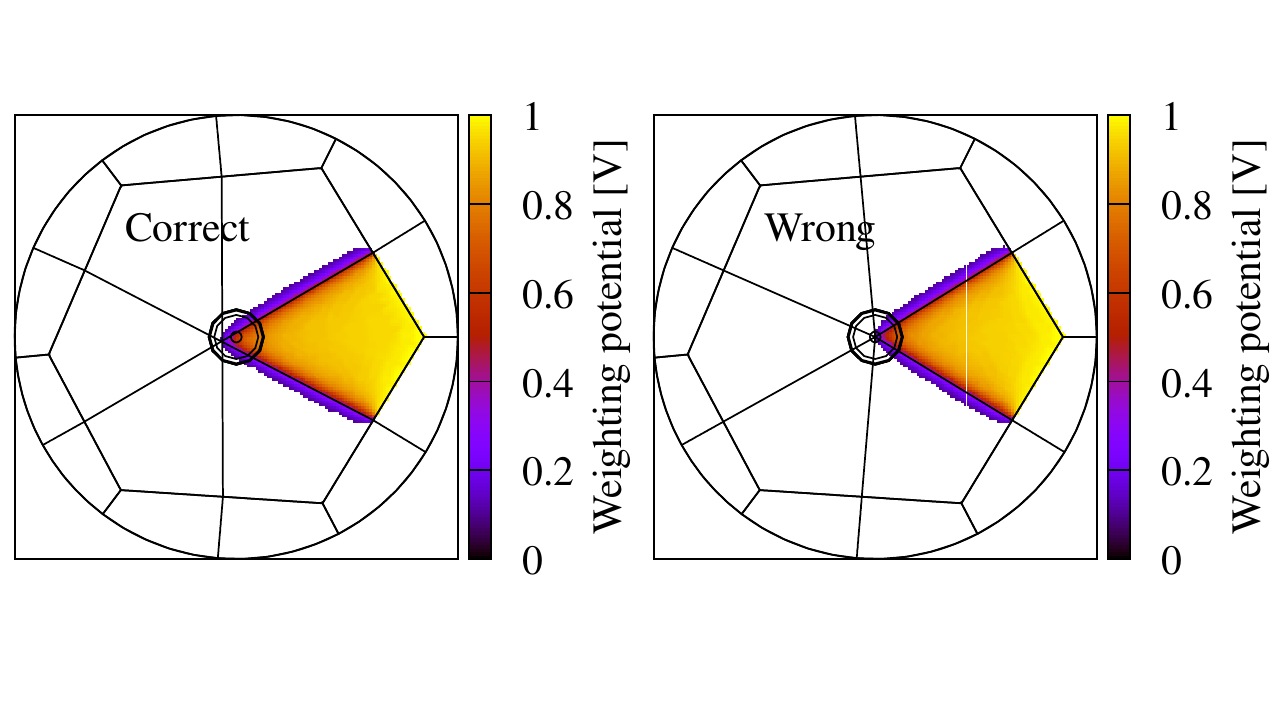}
\caption{\label{fig:org7519ef3}
The difference between the correct (left) and "naive" (right) segmentation.}
\end{figure}

\begin{figure}[htbp]
\centering
\includegraphics[width=.9\linewidth]{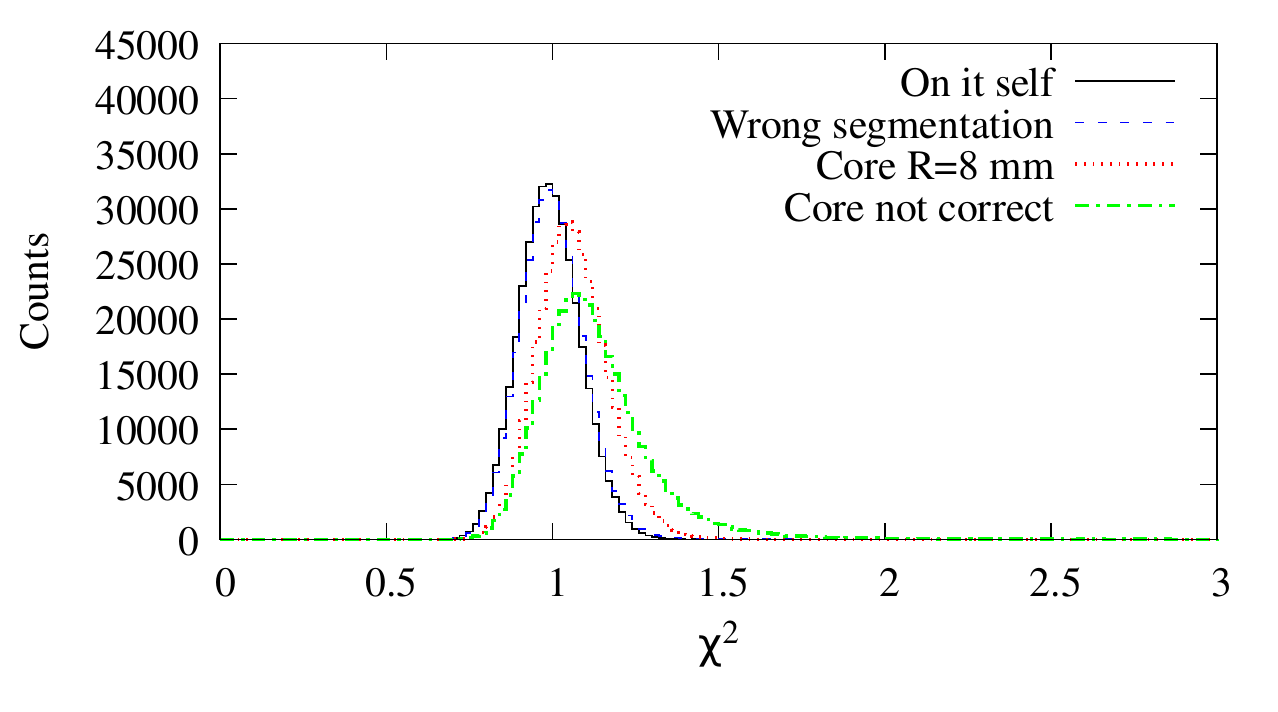}
\caption{\label{fig:org24c409c}
Chi-square distributions for the four different cases of PSA using one exact and three inexact geometries. The \(\chi^2\) rest close to 1 even if the basis used is calculated for a geometry that does not coincide with the correct one.}
\end{figure}

Using the nominal geometry of an A type crystal with a
representative impurity concentration a pulse-shape basis for PSA
with a grid size of 1x1x1 mm\(^3\) and a sample rate of 100 MHz was
first calculated. Using three differently modified geometries the
same 1x1x1 mm\(^3\) grid of points were used to calculate pulse
shapes for each of those. The following geometry modifications were
used: 1) An incorrect front-face segmentation. 2) A displaced
and tilted borehole for the central contact. 3) An enlarged bore hole
for the central contact. The nominal geometry and geometries 2 and 3
are shown in figure \ref{fig:orgf582c04} whereas the difference in front
segmentation is shown in figure \ref{fig:org7519ef3}. The pulse-shape
basis of the nominal geometry was then used for PSA on the
three different pulse-shape sets corresponding to each geometry
variation and on itself as a reference. The used PSA is a simple
extensive grid search with 5 keV Gaussian noise added and an
interaction energy of 1 MeV. In figure \ref{fig:org24c409c} the \(\chi^2\) for the
four different cases are shown. It can be seen that for none of the
geometries it is possible to use the \(\chi^2\) distribution to make
a statement on whether the geometry used to produce the basis is
good or bad - all four distributions are reasonable.

\begin{figure}[htbp]
\centering
\includegraphics[width=.9\linewidth]{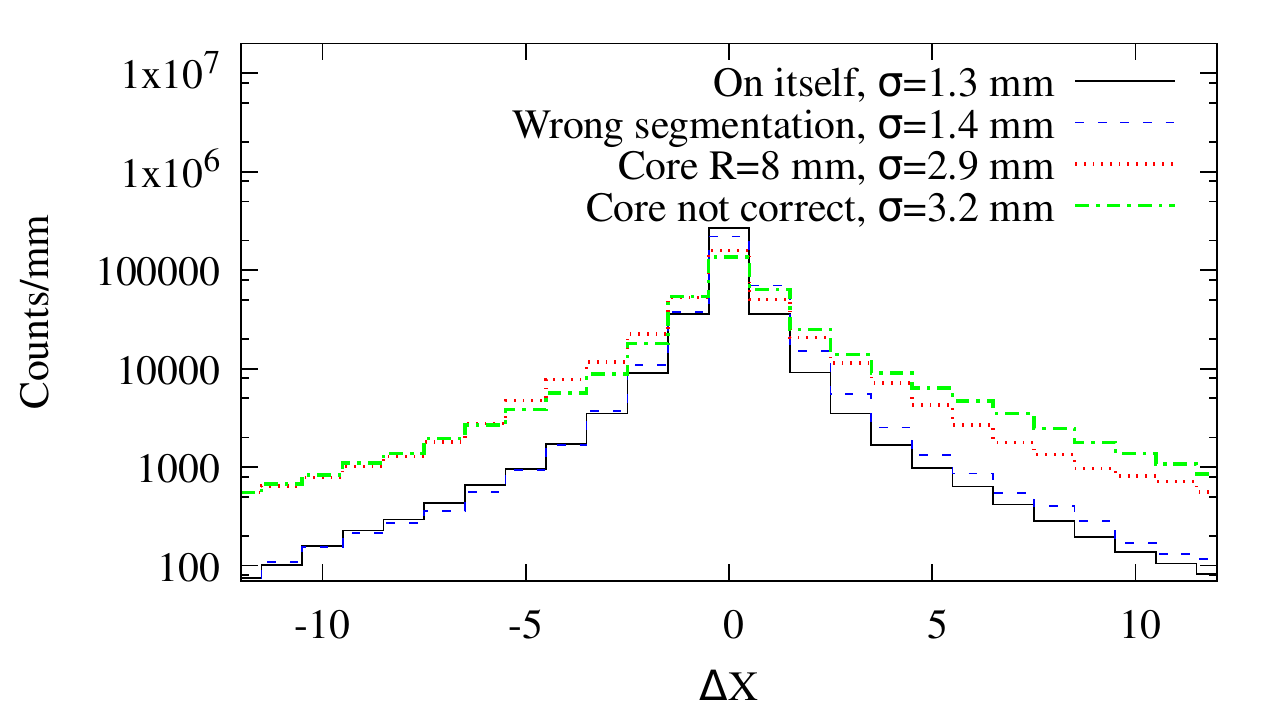}
\caption{\label{fig:orgde58bc4}
The difference between the position given by PSA as compared to the position in the crystal for the calculated signal. This for the X coordinate. The other coordinates are similar. In all four cases the PSA used the "nominal" geometry basis. For details see text.}
\end{figure}

The average error on the determined positions (figure \ref{fig:orgde58bc4} shows
this for the x coordinate) is the most important parameter
evaluating the performance of a pulse-shape basis. The incorrect
segmentation lines do not produce an error that is significant as
compared to the experimentally determined value (\(\sigma\)\textasciitilde{}1.7 mm
\cite{Soderstrom201196}). The two other geometries give an error in
the determination of the actual position that is larger than the
experimental result. A tentative conclusion is that the geometries
of actual AGATA crystals are better known than the two rather
extreme cases used for this test.

\begin{figure}[htbp]
\centering
\includegraphics[width=.9\linewidth]{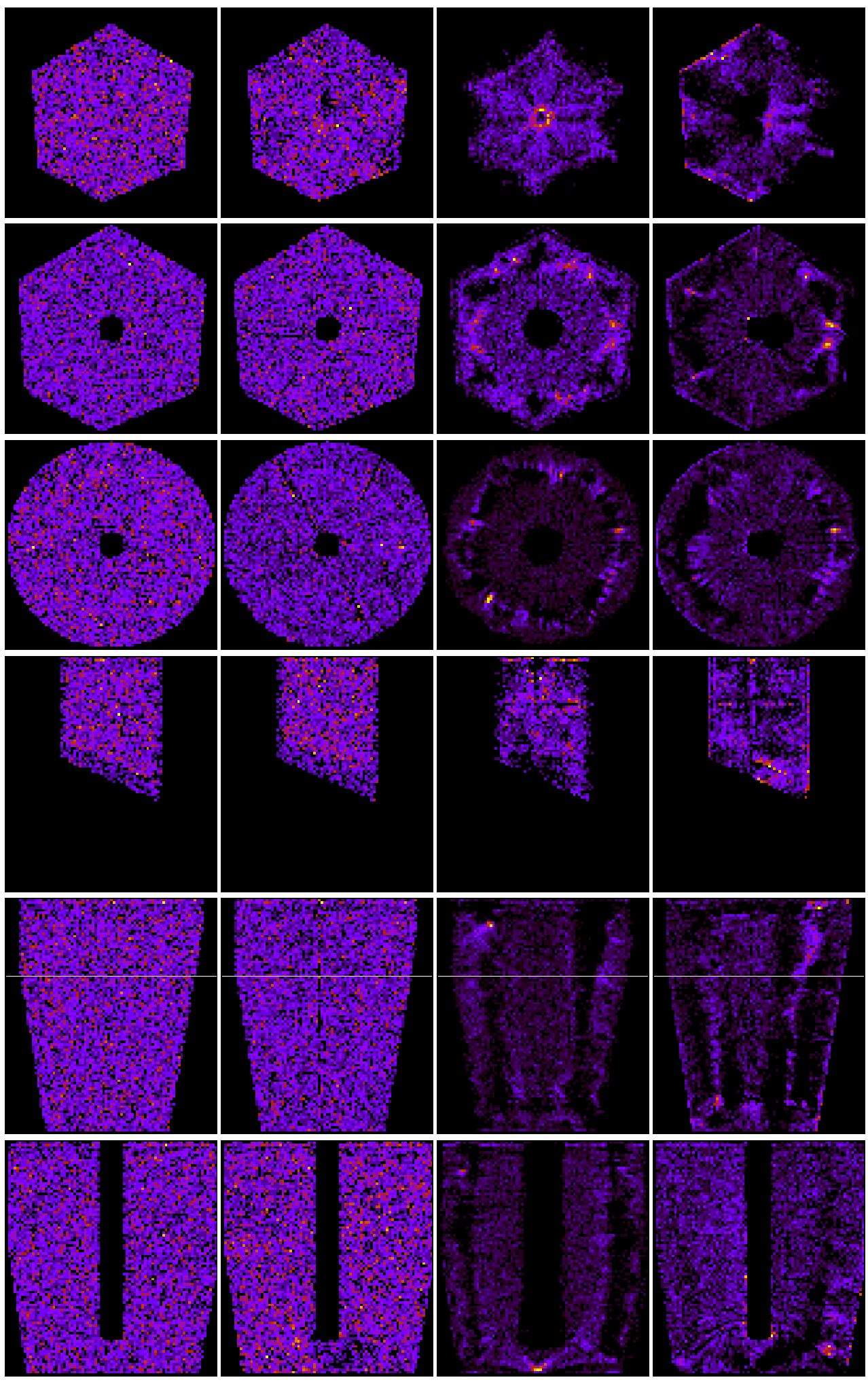}
\caption{\label{fig:orgfaf4d13}
Scatter plots of positions determined with PSA using a basis calculated with the nominal AGATA A type detector. From left, PSA performed with the basis on itself with noise added, PSA performed on signals calculated with an incorrect front-face segmentation, PSA performed with signals calculated using a too large central contact diameter, and finally, to the extreme right, PSA performed using signals calculated with a bore hole for the central contact off center and tilted.}
\end{figure}

Looking at scatter plots of the determined interaction positions,
see figure \ref{fig:orgfaf4d13}, no clustering effects are seen for the
case where PSA is done on pulses belonging to the reference basis
(i.e. on itself) nor with a small error on the front segmentation.
However, when the basis is made with a geometry that deviates from
the geometry of the detector, one of the effects is clustering of
events. The origin of this clustering is that the rise times contain
most of the information in the pulse shapes and all pulses with
extreme rise times will be clustered towards the position in the
basis with the closest rise time. The empty voids are a combination
of this rise-time mismatch and the union of the central contacts of
the nominal geometry and that of the two variations of the bore hole
geometry.

\section{Evaluation of the impact on pulse-shape analyses of crosstalk and noise}
\label{sec:orga1fe921}
\label{org78194c8} Using the AGATAGeFEM package the resolution for
Extensive Grid Search (EGS) and the Singular Value Decomposition
matrix inversion (SVD) \cite{Desesquelles2009} as a function of noise
level and the inclusion of differential and linear crosstalk has
been investigated. The amount of linear crosstalk used for this
investigation is typical for AGATA crystals when mounted in AGATA
triple cryostats \cite{Bruyneel2009196,BRUYNEEL200999} and is about
one per mil. The differential crosstalk is assumed proportional to
the linear crosstalk with the proportionality factor taken as the
one used in the AGATA online PSA (a factor of 10). 

\begin{figure}[htb]
\centering
\includegraphics[angle=0,width=0.5\textwidth]{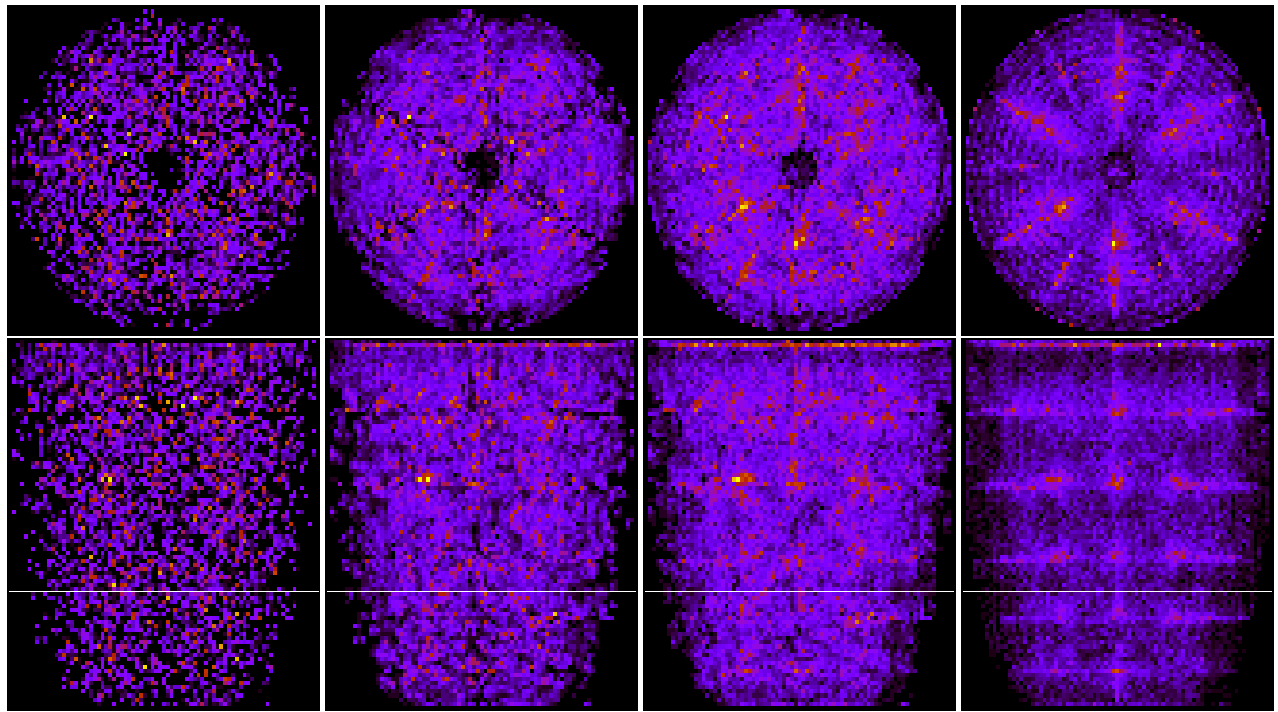}
\caption{\label{fig:org8a5bc51}
Two dimensional projections of positions determined by EGS on calculated pulse shapes. The level of noise have been varied in the interval .6\%→12\%. Note clustering close to segment boundaries for low signal-to-noise ratio.}
\end{figure}

\begin{figure}[htb]
\centering
\includegraphics[angle=0,width=0.5\textwidth]{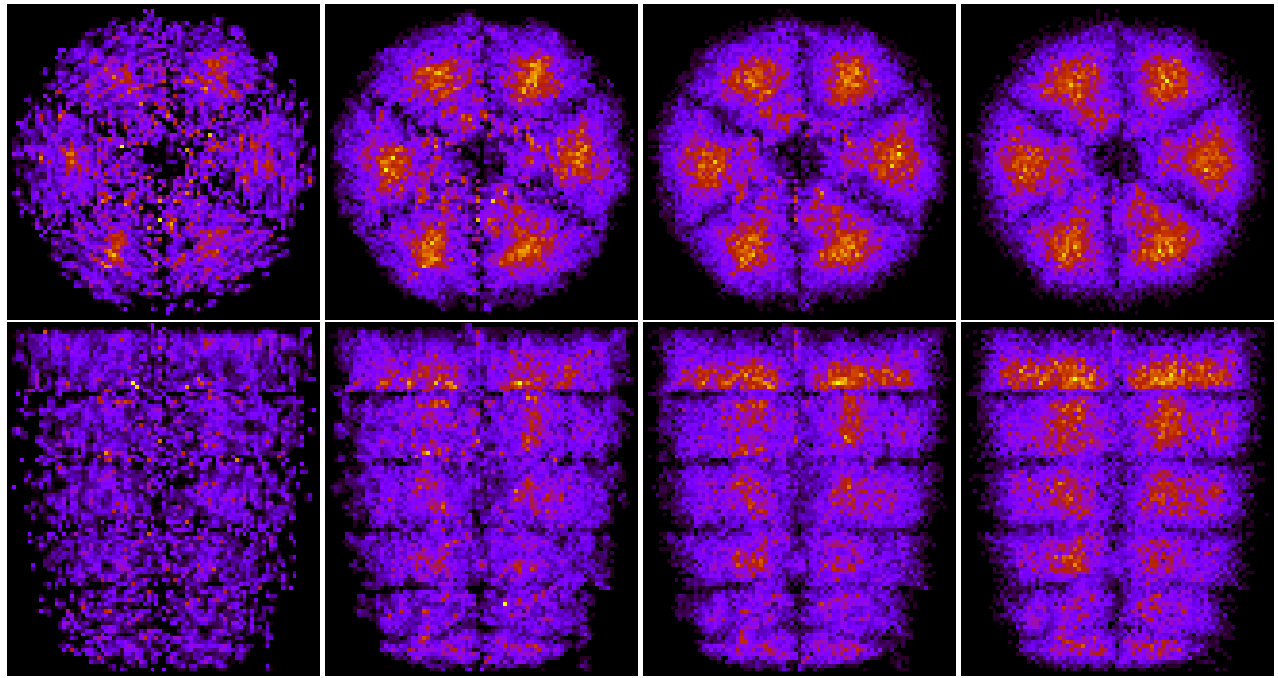}
\caption{\label{fig:org87182c5}
Two dimensional projections of positions determined by SVD on calculated pulse shapes. The level of noise have been varied in the interval .6\%→12\%. Note how the larger noise drives the results towards the center of the segments.}
\end{figure}

Assuming that the physics of a segmented germanium detector is well
known, the possibility to determine the coordinates of a
\(\gamma\)-ray interaction in a large volume HPGe detector is
limited by the knowledge of the response of the electronics and by
the signal-to-noise ratio. These two aspects have been studied by
performing PSA on a data set of pulse shapes calculated using the
AGATAGeFEM with exactly the same parameters as the basis used by the
PSA code. Each pulse shape in the data set was analysed using 6
different levels of noise and using EGS and SVD. This with or
without linear and differential crosstalk for a total of 48
different combinations. Each pulse was analyzed 20 times with noise
regenerated for each time. For both PSA methods linear and
differential crosstalks were added to the analyzed pulses but not to
the pulse-shape basis used for the PSA. The results are summarized
in table \ref{tab:org59682af}. According to this work the crosstalks have a very
limited influence on the resolution, both for the average
reconstructed position and by not introducing systematic errors. In
figures \ref{fig:org8a5bc51} and \ref{fig:org87182c5} two-dimensional projections
of interaction positions as determined by the two different PSA
algorithms are shown. Looking at figure \ref{fig:org8a5bc51} and
\ref{fig:org87182c5} a striking difference shows up for large noises. The
EGS tends to cluster points towards the segment boundaries whereas
the SVD seems to move the points towards the barycenter of the
segment.

\begin{figure}[htb]
\centering
\includegraphics[angle=0,width=0.5\textwidth]{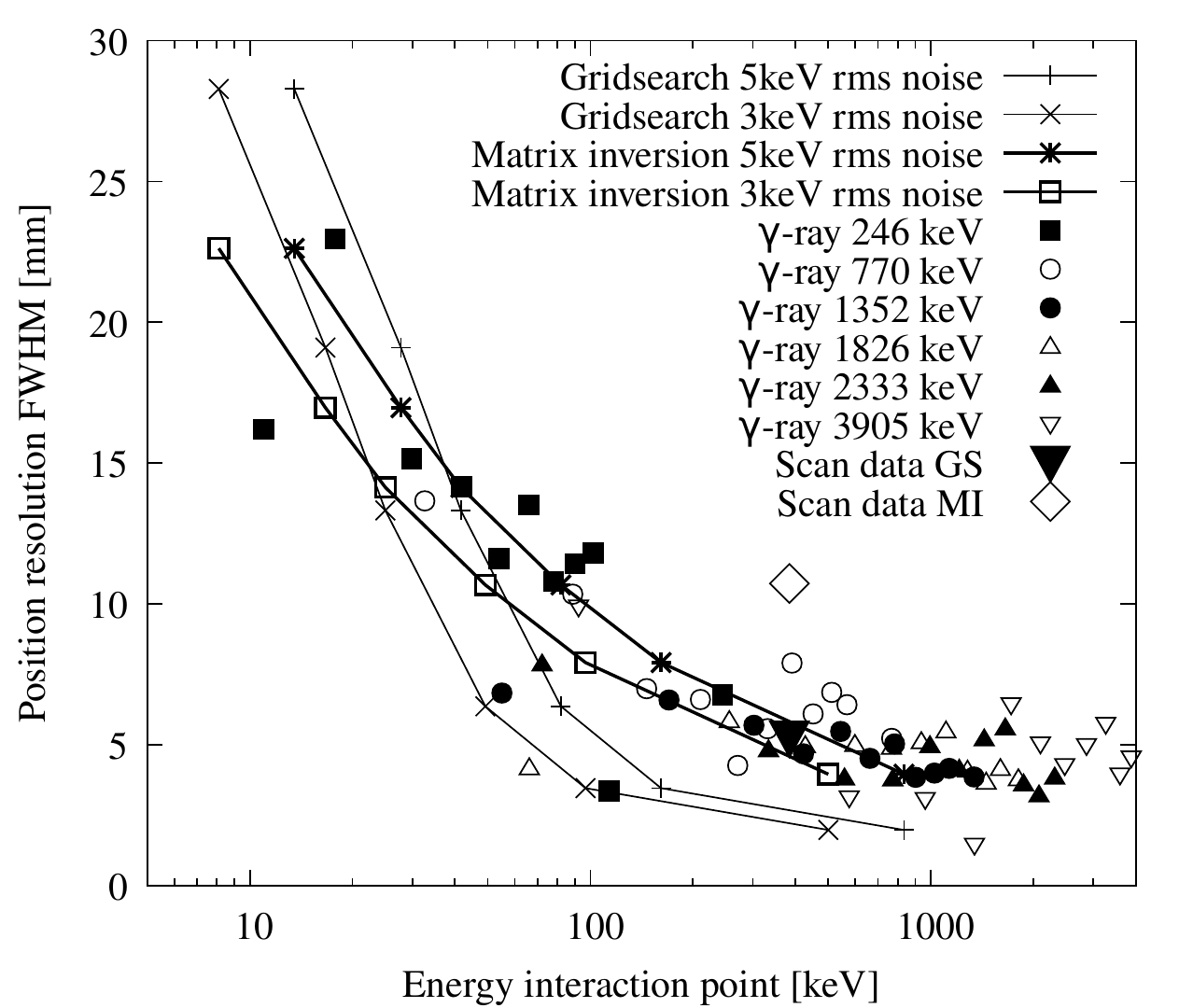}
\caption{\label{fig:org48cee75}
Position resolution as a function of \(\gamma\)-ray interaction energy for simulated data, for different \(\gamma\)-ray energies \cite{Soderstrom201196}, and for data from the 3D-scanning of the S002 at Liverpool \cite{Dimmock2009,Dimmock20091} using extensive grid search (GS) or SVD (MI).}
\end{figure}

Data from Söderström et al. \cite{Soderstrom201196} is presented
together with the result from present work in figure \ref{fig:org48cee75}. One
notes that the EGS is better on simulated data than what has been
experimentally measured for energies above about 50 keV. The Matrix
inversion using SVD decomposition to increase the signal-to-noise
ratio is better at very low interaction energies. It is of interest
to try SVD on low-energy experimental data.

\begin{table}[htbp]
\caption{\label{tab:org59682af}
Full width at half maximum of the distributions interaction positions for different levels of noise for EGS and SVD PSA on a 1x1x1 mm\(^3\) basis.}
\centering
\begin{tabular}{|c|c|c|c|}
\hline
 & No crosstalk &  & \\
\hline
Noise &  & GS [mm] & \\
\hline
[\% rms] & \(\Delta\)x & \(\Delta\)y & \(\Delta\)z\\
\hline
0.6 & 1.3 & 1.4 & 1.3\\
3.1 & 2.4 & 2.5 & 2.4\\
6.1 & 4.2 & 4.7 & 4.2\\
12 & 8.8 & 10 & 8.9\\
18 & 13 & 14 & 13\\
37 & 20 & 19 & 17\\
\hline
 &  & SVD [mm] & \\
\hline
 & \(\Delta\)x & \(\Delta\)y & \(\Delta\)z\\
\hline
0.6 & 2.4 & 2.5 & 1.8\\
3.1 & 5.5 & 5.7 & 4.2\\
6.1 & 7.5 & 7.8 & 5.9\\
12 & 10 & 10 & 8.4\\
18 & 12 & 12 & 10\\
37 & 16 & 16 & 13\\
\hline
 & Full crosstalk &  & \\
\hline
Noise &  & GS   [mm] & \\
\hline
[\% rms] & \(\Delta\)x & \(\Delta\)y & \(\Delta\)z\\
\hline
0.6 & 1.4 & 1.4 & 1.3\\
3.1 & 2.4 & 2.5 & 2.4\\
6.1 & 4.3 & 4.7 & 4.2\\
12 & 8.8 & 10 & 8.9\\
18 & 13 & 14 & 13\\
37 & 20 & 20 & 17\\
\hline
 &  & SVD  [mm] & \\
\hline
 & \(\Delta\)x & \(\Delta\)y & \(\Delta\)z\\
\hline
0.6 & 2.8 & 2.8 & 2.0\\
3.1 & 5.5 & 5.7 & 4.3\\
6.1 & 7.4 & 7.7 & 5.9\\
12 & 10 & 10 & 8.4\\
18 & 12 & 12 & 10\\
37 & 16 & 16 & 13\\
\hline
\end{tabular}
\end{table}

\begin{figure}[htb]
\centering
\includegraphics[angle=0,width=0.5\textwidth]{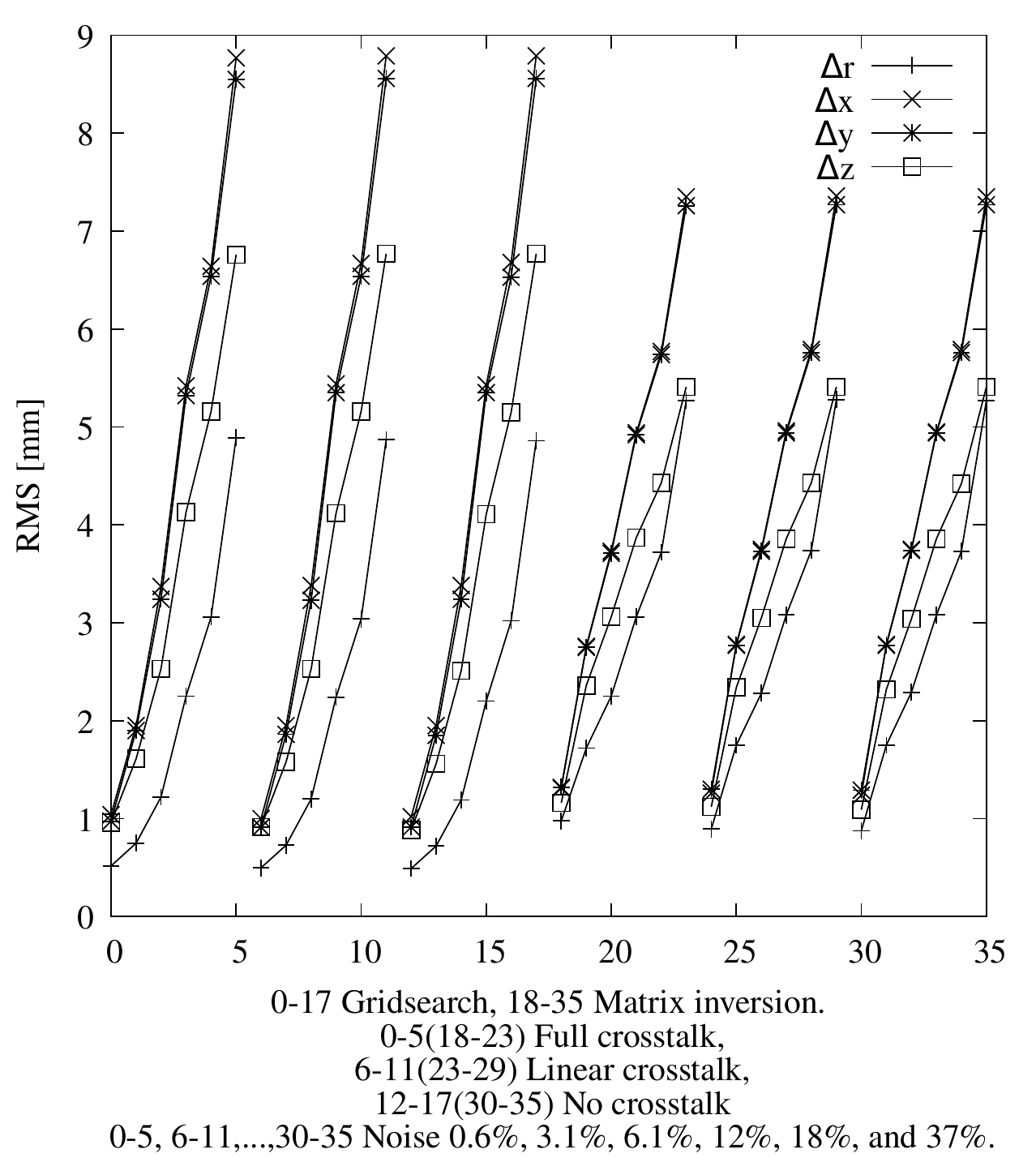}
\caption{\label{fig:org3c92039}
RMS values for different amounts of noise for EGS and SVD PSA on a 1x1x1 mm\(^3\) basis. Results without crosstalk, with only linear crosstalk, or with linear and differential crosstalk added to the test signals are shown.}
\end{figure}

\section{Pulse-shape analysis on experimental data}
\label{sec:org10b0b64}
\label{org780aa49} One of the objectives of the AGATAGeFEM package is
to produce pulse-shapes bases used for the PSA of experimental
data. In the AGATA collaboration an adaptive grid search algorithm is
presently used \cite{ventrurelli2004}. The validation of bases
calculated with AGATAGeFEM for the AGATA PSA is presented
in this section. This step also validates the AGATAGeFEM package for
use in the development of pulse-shape analyses by proving that the
pulse shapes are realistic. Pulse-shape data bases have been
calculated for 6 AGATA crystals. These crystals were previously used
to estimate the achieved position resolution
\cite{Li2018,LJUNGVALL2020163297} employing bases calculated with ADL
\cite{Bruyneel20161,Bruyneel2016}.

An important parameter when calculating the electric fields inside
the in fully depleted detectors is the space charge. The space
charges come from impurities in the Germanium crystals. For the
AGATA crystals they have been measured by the community using
techniques based on the capacity of the crystal
\cite{BIRKENBACH2011176,BRUYNEEL201192} and are used as input here
with numerical values presented in table \ref{tab:orge809380} in appendix
\ref{org56d91ba}. The parameters for the electron and hole mobility are taken
from Ljungvall et al. \cite{LjungvallThesis}. They differ slightly
from values used in ADL \cite{Bruyneel20161,Bruyneel2016} and are
presented in table \ref{tab:org047b7d6} appendix \ref{org56d91ba}.

The addition of electronics transfer function and crosstalk to the
pulse-shape data bases were performed by the standard AGATA PSA
codes. For the crosstalk values measured during each experiment for
each crystal are used (for typical values see Bruyneel at al.
\cite{Bruyneel2009196,BRUYNEEL200999}). The response function of the
electronics is modeled as an exponential with a rise time of 35 ns,
the default used for PSA in AGATA.

\begin{figure}[htbp]
\centering
\includegraphics[width=.9\linewidth]{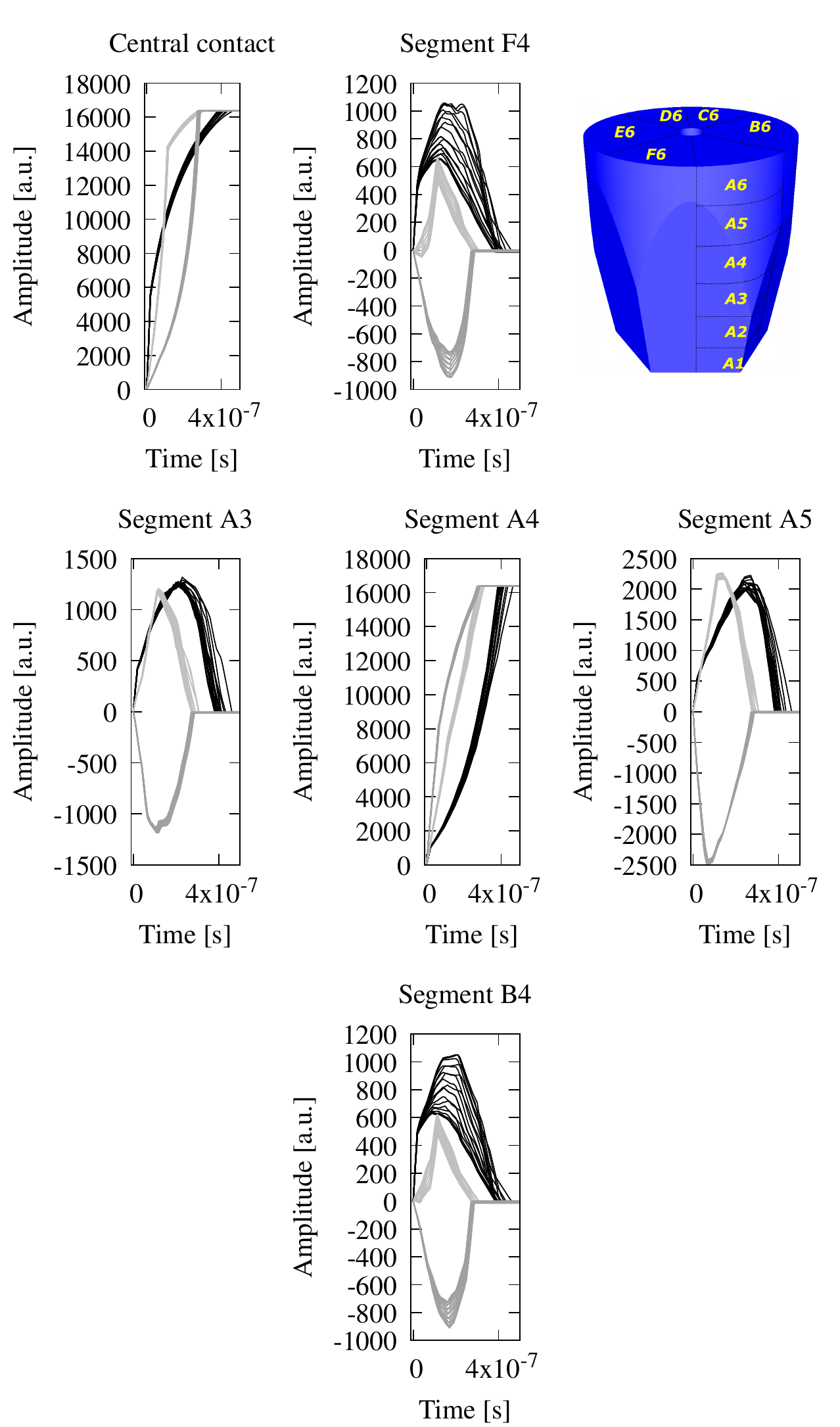}
\caption{\label{fig:orga05dbba}
Variation in rise time and shape of transient signals as a function of crystal lattice angle. The figure shows changes for the pulse shapes as the crystal lattice is rotated over \(90^{\circ}\) for three different radii , R=8 mm (black), 20 mm (grey), and 36 mm (dark grey), respectively.}
\end{figure}

\begin{figure}[htbp]
\centering
\includegraphics[width=.9\linewidth]{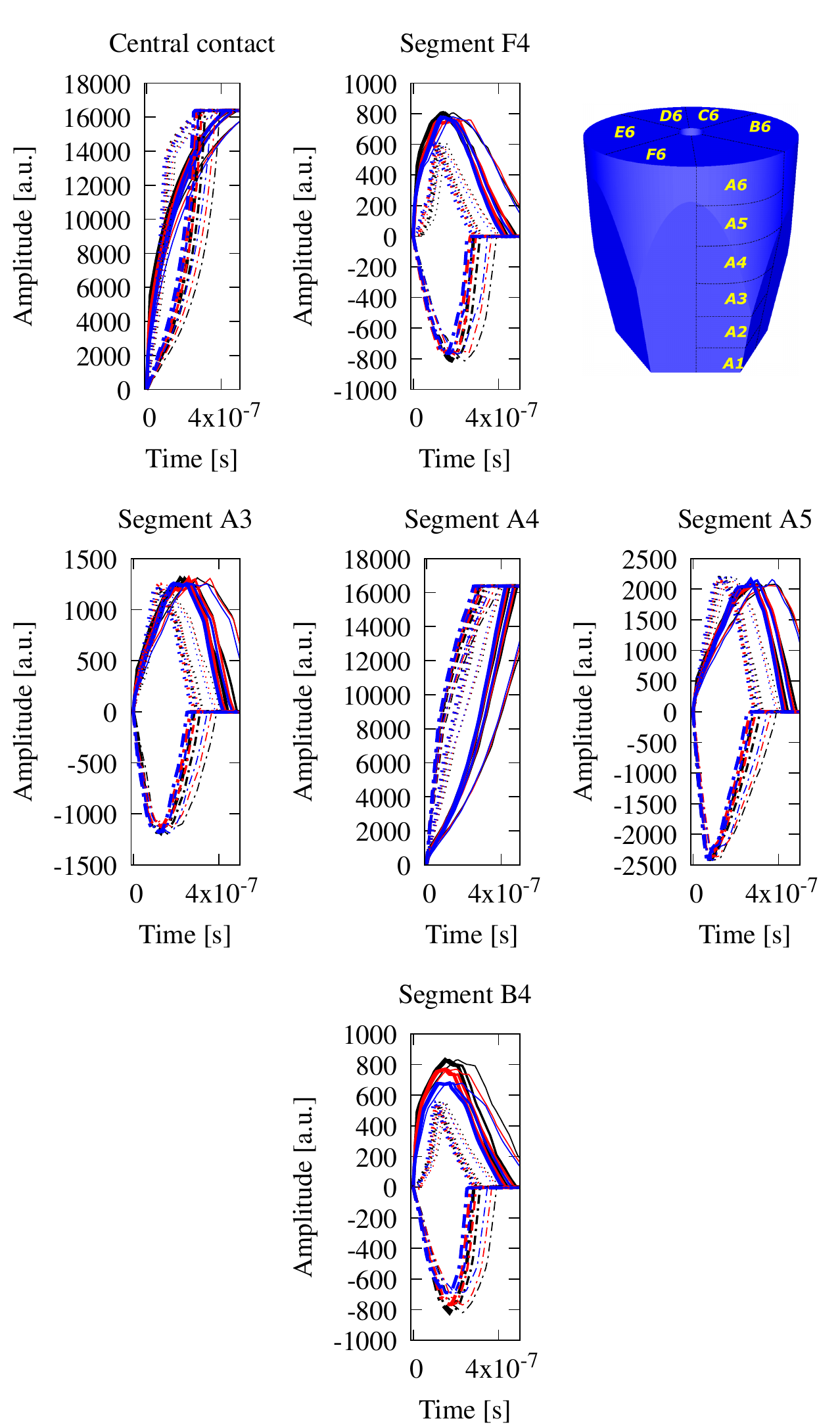}
\caption{\label{fig:orgdf8abe7}
Variation in rise time and shape of transient signals as a function of central contact radius (black=5mm, red=6mm, blue=7mm), and scaled charge-carrier velocities (thin=0.8, thicker=1.0, thickest=1.1), respectively. This is shown for interaction at three different radii (solid=8 mm, dotted=20mm, dashed-dotted=36 mm).}
\end{figure}

\begin{figure}[htbp]
\centering
\includegraphics[width=.9\linewidth]{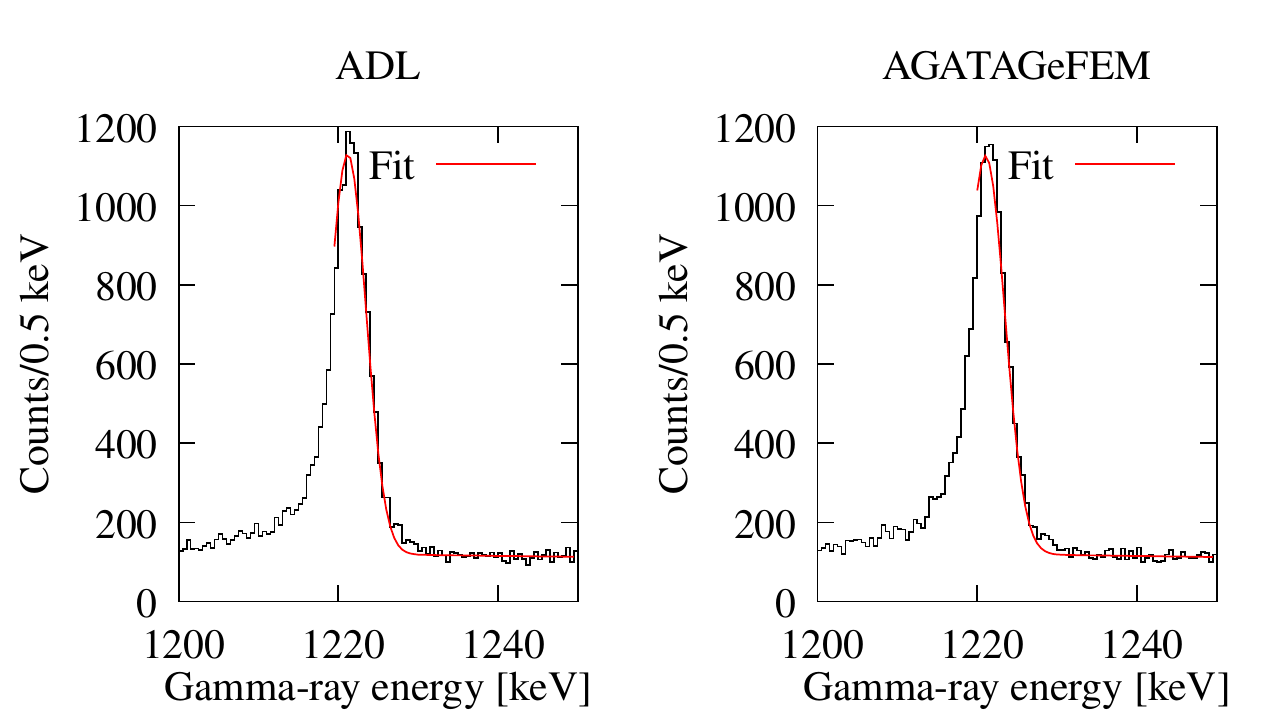}
\caption{\label{fig:org54366c3}
Examples of how the Full Width at Half Maximum is extracted from the data. An automatic procedure has been chosen to minimize biases. The fit has been done on the upper half of the peak because of the tail to the left that is a result of the nucleus slowing down in the target.}
\end{figure}

To optimize the calculated bases, the parameters controlling the
direction of the <100> crystal axis, the assumed radius of the
central contact, and scaling of the hole and electron velocities,
were varied. The best results for the PSA was sought in this
parameter space for each crystal. The orientation of crystal lattice
was varied by rotating the lattice around the <100> crystal axis
assumed parallel to the bore hole for the central contact. Rotations
in steps of 5 degrees until 90 degrees were performed. For the bore
hole, three different radii were used: 5 mm (nominal), 6 mm, and
7mm. The charge carrier velocities were scaled from \(0.8\) to
\(1.1\) is steps of \(0.1\) of their nominal values. In figure
\ref{fig:orga05dbba} and \ref{fig:orgdf8abe7} the variation of the pulse shapes
over the used parameter space is illustrated. As can be seen in
figure \ref{fig:orga05dbba} the lattice orientation has a noticeable impact
on the pulse shapes. The varied parameters with the largest impact
are the 10 \% step scaling of the charge-carrier velocities, clearly
seen in figure \ref{fig:orgdf8abe7}. This is coherent with what was shown
in section \ref{org39df270}. Evaluation of the performance of the PSA was
done using the peak width of the 1221 keV \(\gamma\)-ray transition
in \(^{90}\)Zr, as illustrated in figure \ref{fig:org54366c3}. Doppler
correction was performed using first interaction point as defined by
the \(\gamma\)-ray tracking and the recoil velocity of the nucleus
given by the VAMOS spectrometer (for details see Li et al.
\cite{Li2018}). An automatic fit procedure was chosen to exclude
biases for one or the other set of bases. The experimental data was
processed, with the exception of choice of pulse-shape data bases,
exactly as described in Ljungvall et al. \cite{LJUNGVALL2020163297}.
For the detectors used in this work neutron-damage correction was
not needed. Similar efforts to optimize the results of PSA using ADL
have recently been published by Lewandowski et al.
\cite{Lewandowski2019}. The reader is cushioned that due to strong
correlations between different parameters entering in the
calculation of pulse shapes and in the PSA it is difficult to
compare individual parameters, especially as different figures of
merits are used.

\begin{figure}[htbp]
\centering
\includegraphics[width=.9\linewidth]{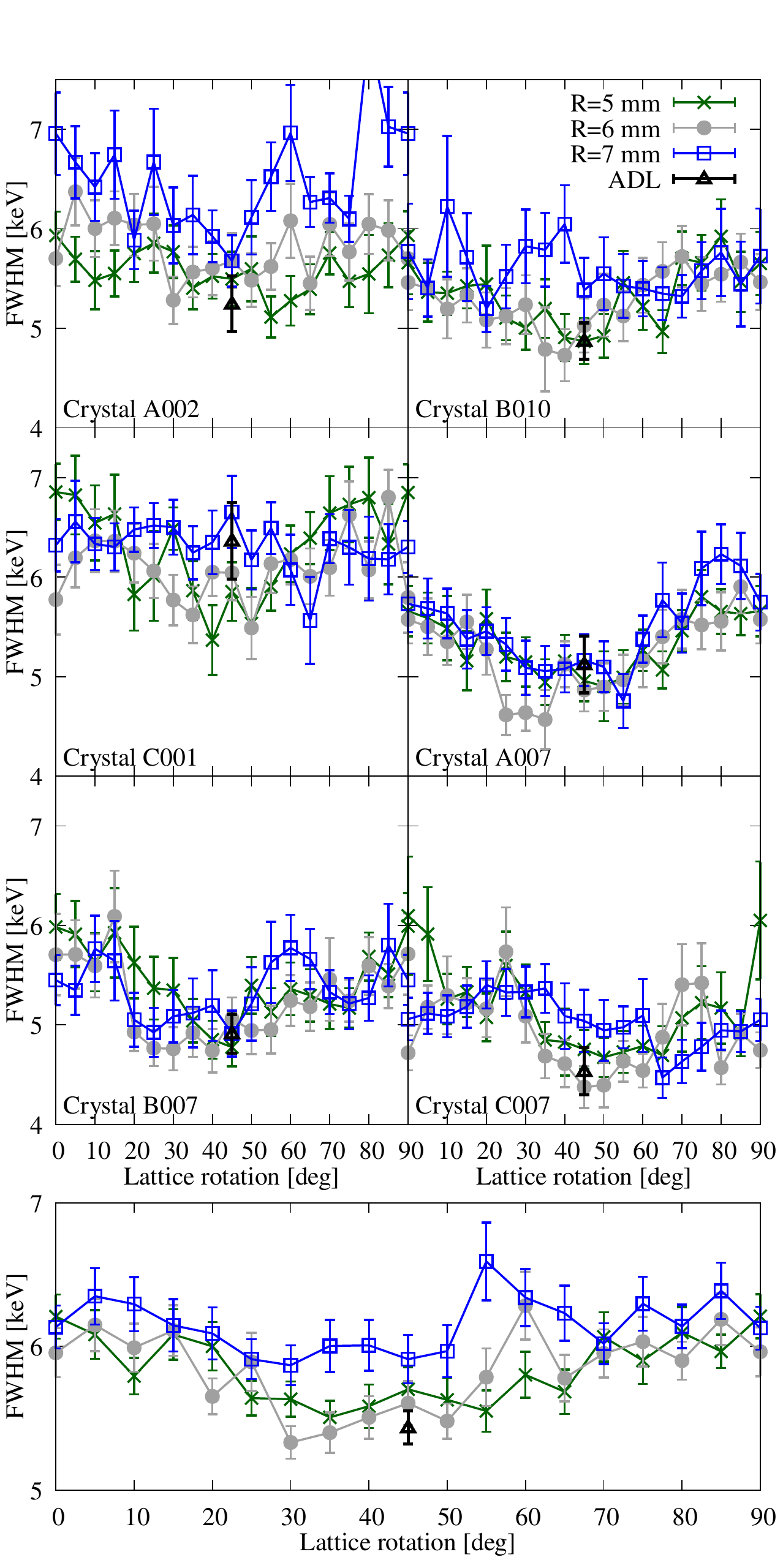}
\caption{\label{fig:orgef4b567}
Extracted Full Width at Half Maximum for the 1221 keV \(\gamma\)-ray peak in \(^{90}\)Sr as a function of lattice rotation and assumed radius on the central contact. Results using the standard AGATA ADL pulse-shape data bases are also shown.}
\end{figure}

A first optimisation varying only the lattice orientation and the
central contact hole radius was performed. Pulse-shape analyses were
done using a total of 72 different bases for each crystal followed
by \(\gamma\)-ray tracking. The resulting spectra were fitted by an
automatic routine and the FWHM were extracted for each crystal and
for the total spectrum. The results, for each crystal and for the
sum of the crystals, are presented in figure \ref{fig:orgef4b567}. The
minimum FWHM is achieved close to the nominal orientation for most
of the crystals. However, the minimum of the sum is close to a
lattice rotation of 30\(^{\circ}\) and with an assumed core radius
of 6 mm. This seems to be driven by crystal A007. The nominal
direction of the lattice is 45\(^{\circ}\). In figure \ref{fig:orgef4b567}
the results obtained with the data bases calculated using ADL are
also shown as a reference.

\begin{figure}[htbp]
\centering
\includegraphics[width=.9\linewidth]{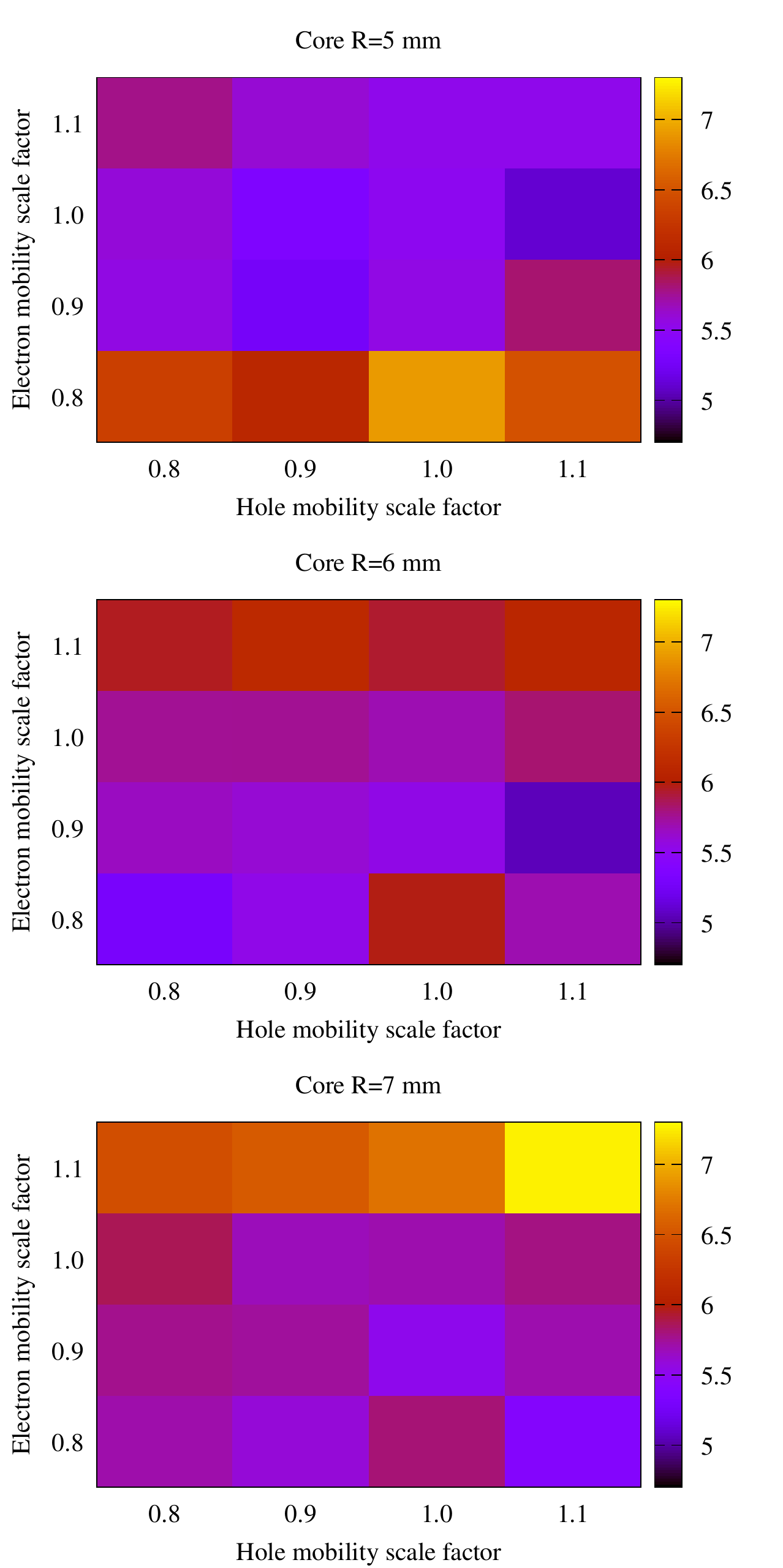}
\caption{\label{fig:org293d06e}
Full Width at Half Maximum of the 1221 keV \(\gamma\)-ray peak in \(^{90}\)Sr for the A002 crystal using pulse-shape bases calculated with different central contact radii and charge-carrier velocity scaling. The crystal lattice was kept fixed to  45\(^{\circ}\).}
\end{figure}

A second minimisation was performed on a parameter space
including the three different central contact radii, 5 different
crystal lattice orientations and 16 different scalings of the charge
carrier velocities. In figure \ref{fig:org293d06e} the variation
of the FWHM for crystal A002 is shown for the three different
central contact radii and as a function of the scale of the electron
and hole velocities. For each data point a crystal lattice orientation
of 45\(^{\circ}\) was chosen. As can be seen there is a correlation
between the central contact radius and the best scale factor for the
electron velocity. It is reasonable that it is the electron velocity
that can be used to compensate for changes in the central contact
radius as they are (mainly) responsible for the generation of the
signal on the central contact. Furthermore, the change of the
central contact radius generates the strongest change in the
electric field in the regions where the largest contribution from the
electrons to the signal is created. From this optimisation it is also
clear that different parameters for the pulse-shape calculations are
strongly correlated and that it is difficult, if not impossible, to
determine individual parameters using only the width of a
\(\gamma\)-ray peak (or any other single figure of merit).

\begin{figure}[htbp]
\centering
\includegraphics[width=.9\linewidth]{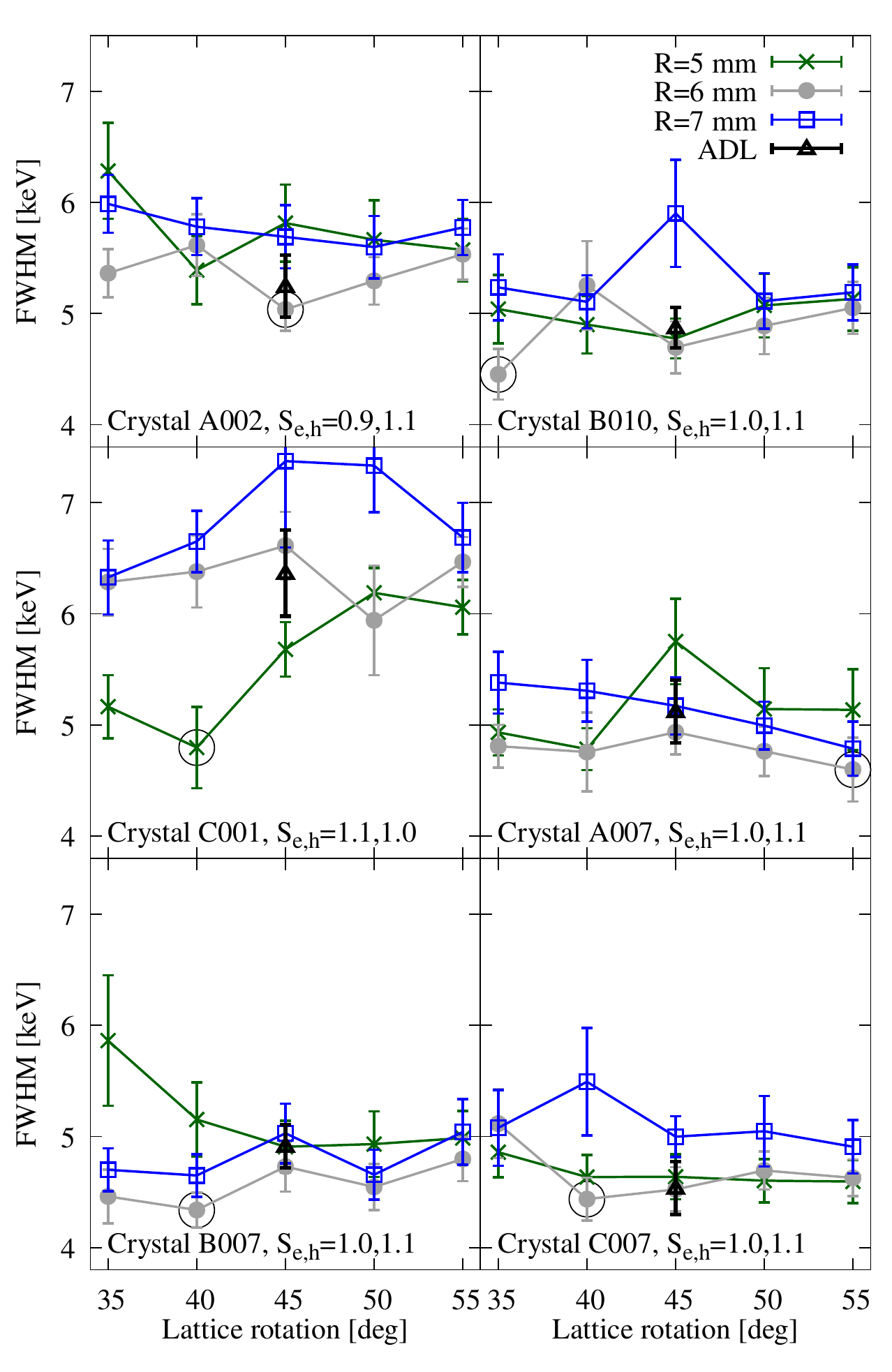}
\caption{\label{fig:org0d39fa9}
Full Width at Half Maximum of the 1221 keV \(\gamma\)-ray peak in \(^{90}\)Sr as a function of crystal lattice orientation for the set of charge carrier velocity scaling giving the smallest FWHM for each crystal used for evaluating the PSA performance. The best combination for each crystal is marked with a circle. Results using the ADL bases are given as reference.}
\end{figure}

From the second minimisation the bases that produce the smallest
FWHM of the 1221 keV peak were chosen for each crystal. In figure
\ref{fig:org0d39fa9} the lowest FWHM for each crystal is marked with a circle.
It seems as if a slight increase of the hole mobilities, reflected
by the scaling factor \(S_h\), improves the performance on average.
With an exception of the C001 crystal. This is not a general
statement about hole mobilities when modeling HPGe detectors but
only applies to this work. As for the bore hole radius, 6 mm seems
to be preferred with, again, C001 in disagreement. Possible reasons
for the different behaviour of the C001 crystal will be discussed
later in this section.

\begin{figure}[htbp]
\centering
\includegraphics[width=.9\linewidth]{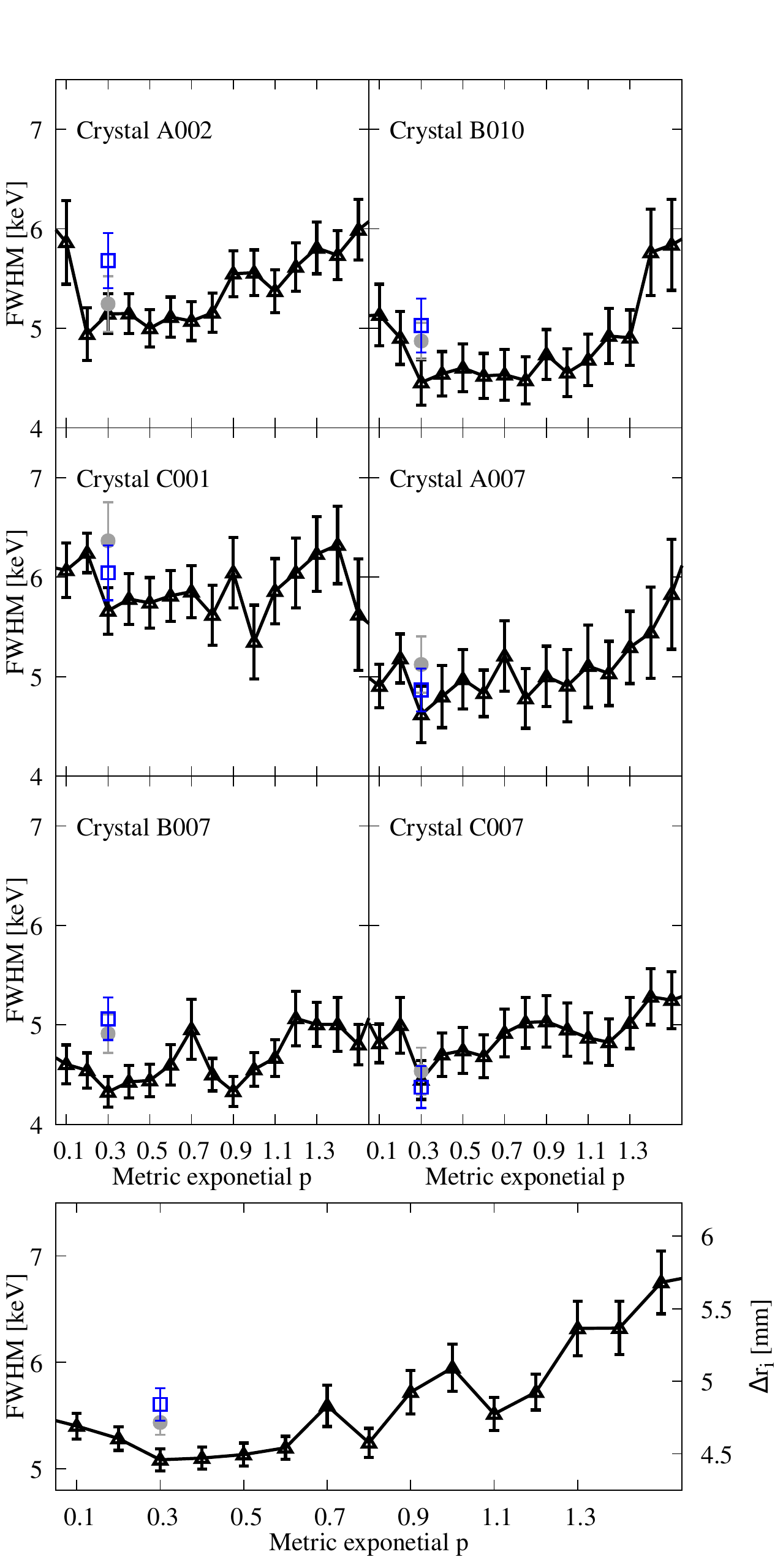}
\caption{\label{fig:org922ac53}
Full Width at Half Maximum of the 1221 keV \(\gamma\)-ray peak in \(^{90}\)Sr as a function of the exponential \(p\)  in the metric used to compare entries in the pulse-shape bases with experimental pulse shapes. The test was performed with the bases selected to be the best according to the FWHM for each crystal. The best value of \(0.3\) is in accordance with earlier results. A range of 0.1-3 in steps of 0.1 was scanned but as the results grow worse as the metric increases beyond \(0.3\) results are only shown up to \(1.5\). Results from using the ADL bases and with AGATAGeFEM bases assuming a central contact radius of 6 mm and a crystal lattice rotation of 45\(^{\circ}\) are given as reference (shown as grey circles and blue rectangles, respectively). On the lower panel an approximate conversion to position resolution (FWHM) is given (see \cite{Li2018}).}
\end{figure}

\begin{figure}[htbp]
\centering
\includegraphics[width=.9\linewidth]{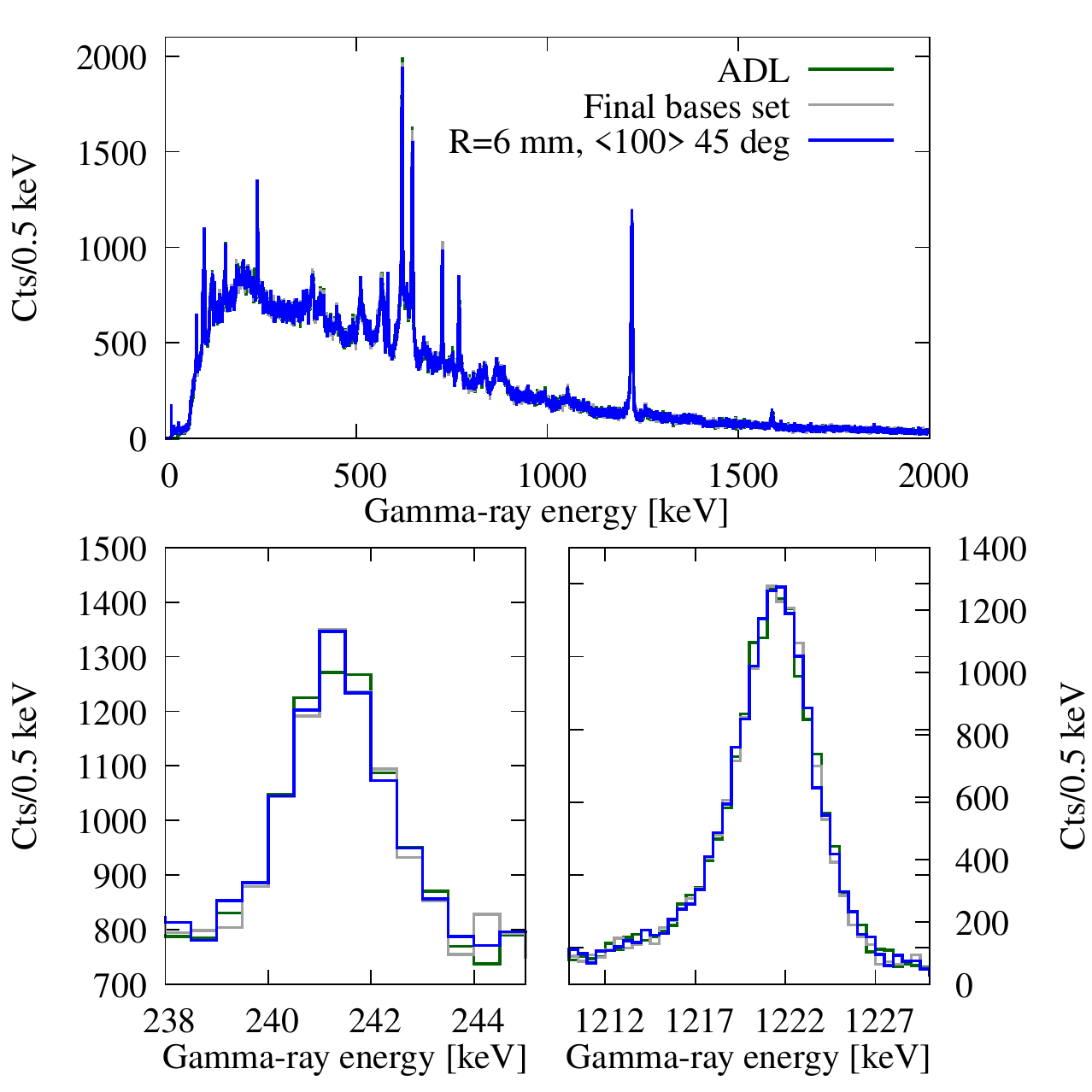}
\caption{\label{fig:org06db098}
Histograms made using three different set of pulse-shape data bases, the standard ADL data base, an AGATAGeFEM data base, and the optimized AGATAGeFEM data base.}
\end{figure}

The most important parameter in the adaptive grid search PSA
algorithm used within the AGATA collaboration \cite{ventrurelli2004}
is the power used to calculate the figure of merit (FOM) for a
pulse shape in the basis when compared to the experimental signal.
A minimum is sought for the expression 
\begin{eqnarray}
\label{eq:org1a3a0c7}
  \sum_i|y_{i}^{exp}-y_{i}^{base}|^p
\end{eqnarray}
which for \(p=2\) is the typical square sum FOM and i is the index
of the samples points. Using the ADL bases it has been shown
that an \(p\) of \(0.3\) give the best result \cite{RecchiPHD} (for a
recent in-depth discussion of the impact of the distance metric on
PSA, see Lewandowski et al. \cite{Lewandowski2019}). A scan of \(p\)'s
were performed using the six selected AGATAGeFEM bases, and the
resulting FWHM are presented in figure \ref{fig:org922ac53}. For
reference the FWHM achieved using the ADL bases and the AGATAGeFEM
are also shown. The AGATAGeFEM bases were calculated with a central
contact radius of 6 mm and a lattice orientation of 45\(^{\circ}\).
This is the final result of the optimisations made in this work. On
the lower panel in figure \ref{fig:org922ac53} an approximate conversion
to position resolution has been added (for details see Li et al.
\cite{Li2018}) allowing to estimate the improvements made. Looking
at the lower panel in figure \ref{fig:org922ac53} an improved position
resolution of a few tens of millimeters is suggested, this by
comparing the points using the ADL bases and the optimised
AGATAGeFEM bases (at 0.3 on the x-axis). The corresponding
\(\gamma\)-ray spectra are shown in figure \ref{fig:org06db098}. Visually
the difference in FWHM of 0.35 keV for the peak 1221 keV peak is not
obvious, and no manual fit has been performed to verify the
difference. This in the spirit of minimizing biases when analysing the
data. One can however state that the AGATAGeFEM produced bases
perform as well as the ADL bases presently used for analysing AGATA
data.

\begin{figure}[htbp]
\centering
\includegraphics[width=.9\linewidth]{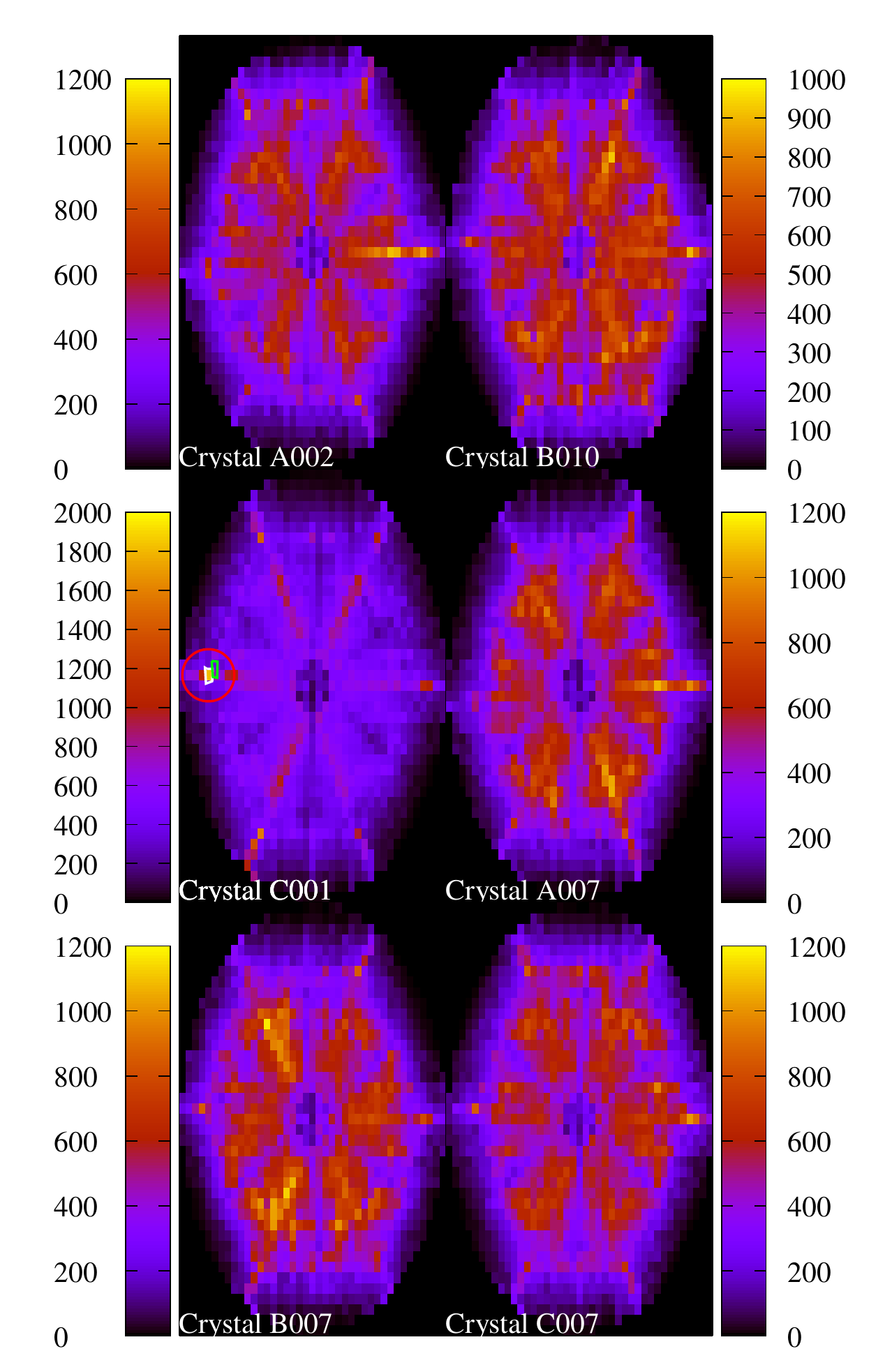}
\caption{\label{fig:org92809e3}
Gamma-ray interaction points as determined with the ADL pulse-shape data base. For crystal C001 the regions used to create averaged traces when investigating the origins of "hot spots" are found inside the red circle and are marked in white for events belonging to the "hot spot" and in green for reference events, respectively (for details see text).}
\end{figure}

\begin{figure}[htbp]
\centering
\includegraphics[width=.9\linewidth]{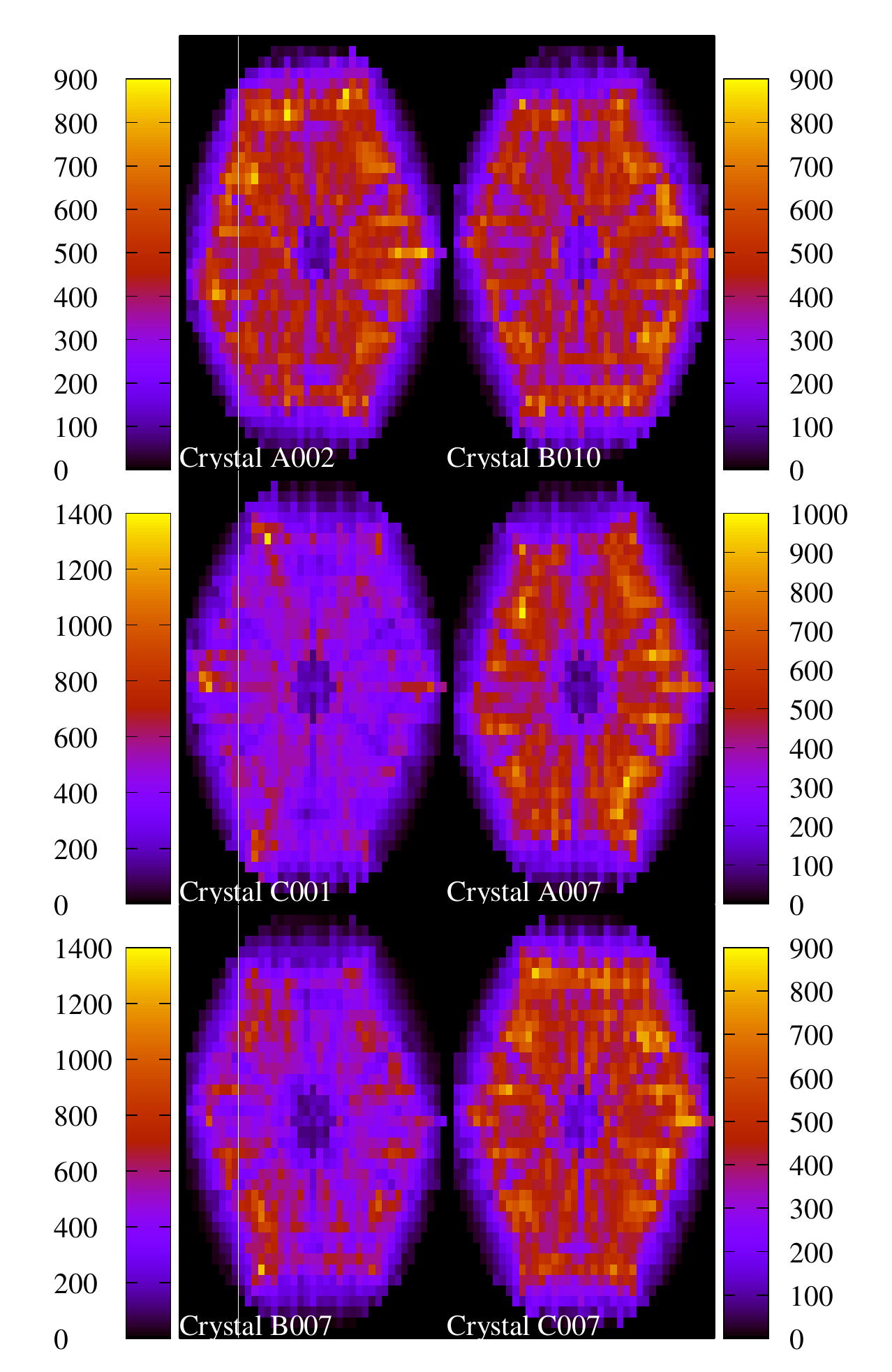}
\caption{\label{fig:org018dac2}
Gamma-ray interaction points as determined with the AGATAGeFEM pulse-shape data base with a central contact hole radius of 6 mm.}
\end{figure}

\begin{figure}[htbp]
\centering
\includegraphics[width=.9\linewidth]{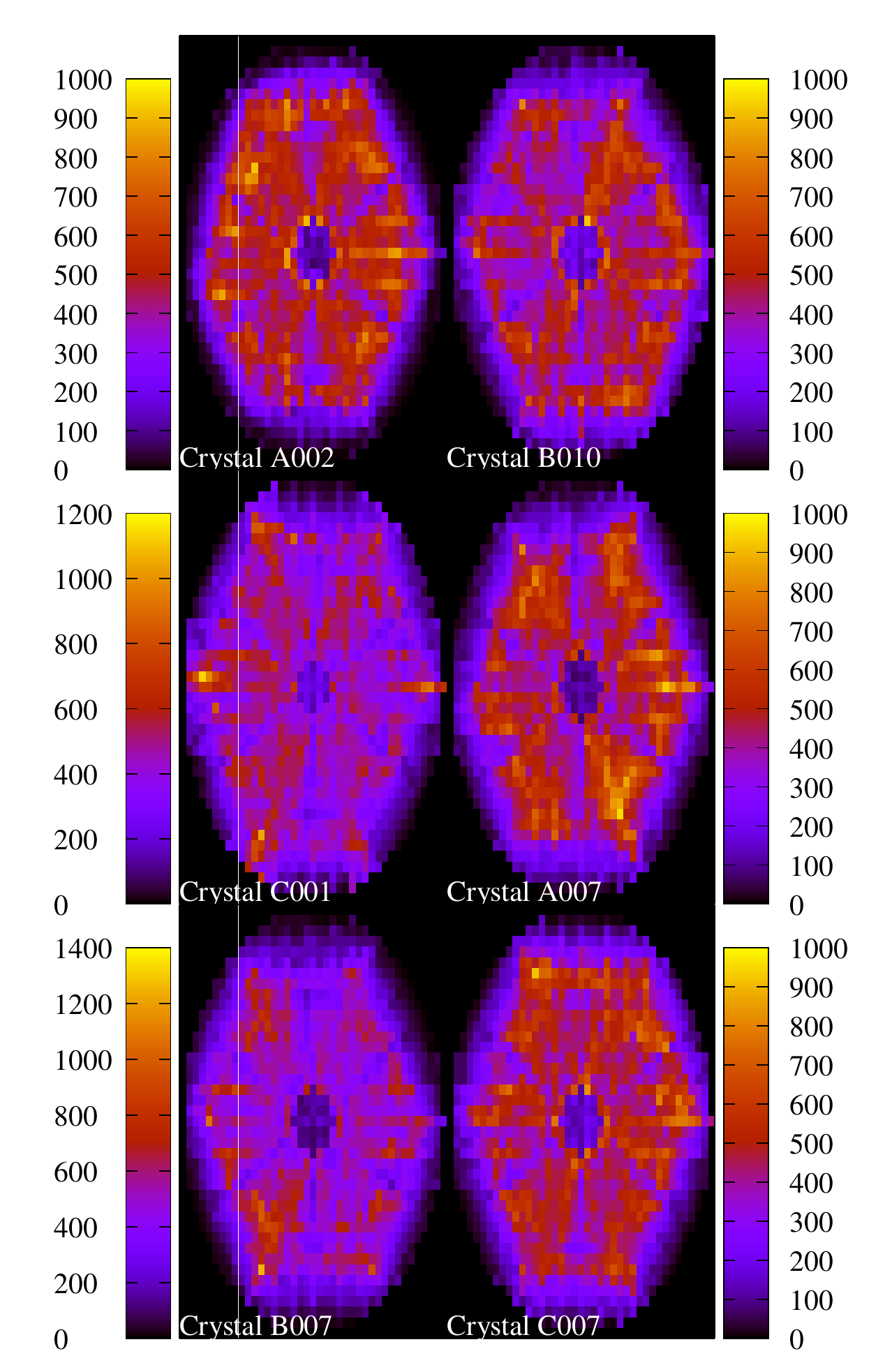}
\caption{\label{fig:org602c4f7}
Gamma-ray interaction points as determined with the best performing  AGATAGeFEM pulse-shape data base.}
\end{figure}

\begin{figure}[htbp]
\centering
\includegraphics[width=.9\linewidth]{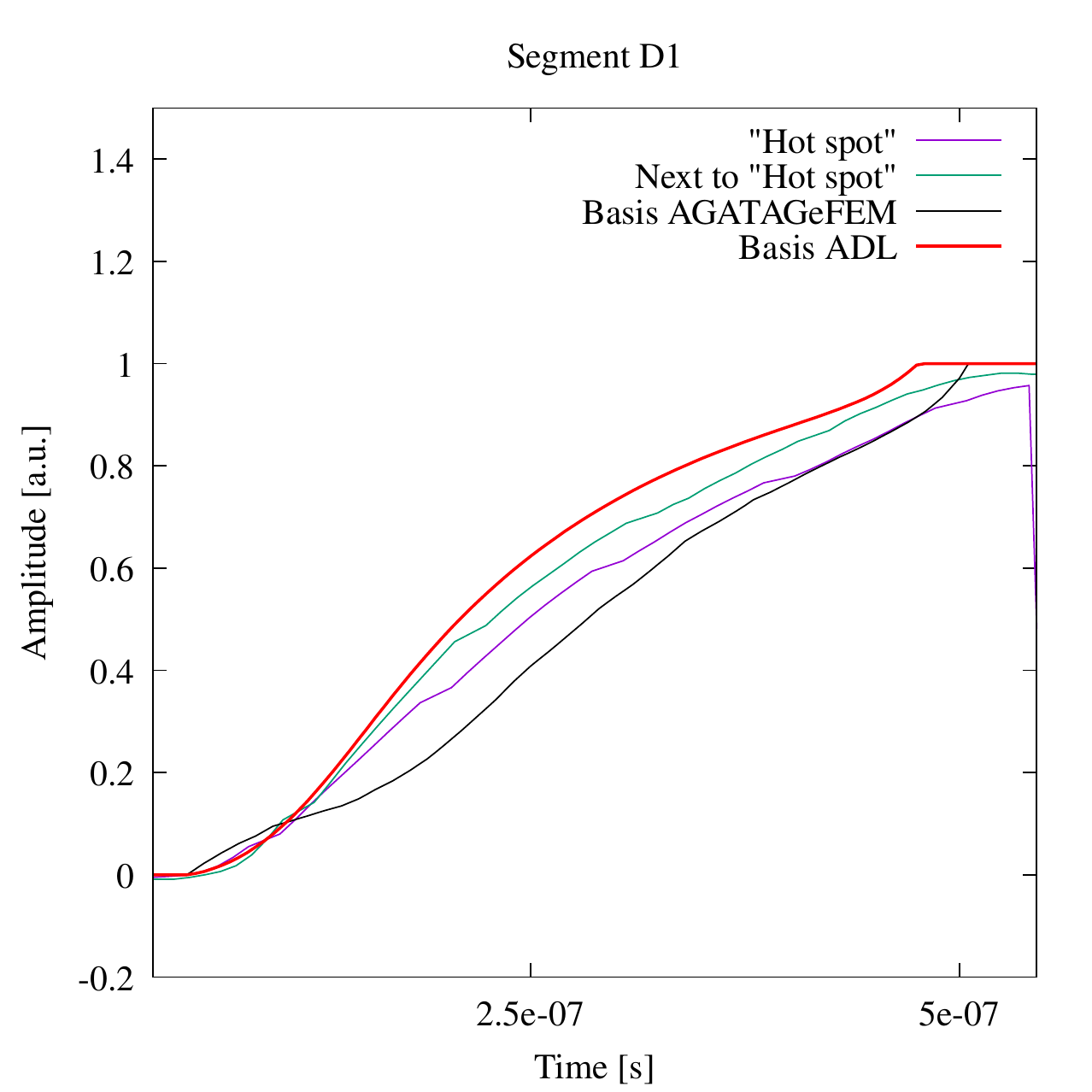}
\caption{\label{fig:orgecd679f}
Signals from the net-charge segment for one-segment events coming from the "hot spot" seen for crystal C001 (for the ADL bases, see figure \ref{fig:org92809e3}).}
\end{figure}

An homogeneous \(\gamma\)-ray flux over the solid angle covered by
one AGATA crystal is an excellent approximation, and as a consequence
the \(\gamma\)-ray interaction points should be homogeneously
distributed on planes parallel to front face of the detector (here
the small effect coming from that strictly speaking we should refer
to spherical surfaces is ignored). Looking at the projections of
\(\gamma\)-ray interaction points onto the plane parallel to the
front face is therefore an indicator of how well the PSA is
performing. In the figures \ref{fig:org92809e3}, \ref{fig:org018dac2}, and \ref{fig:org602c4f7}
projections of \(\gamma\)-ray interaction points are shown for the
six crystals and three different sets of bases, respectively. In all
figures similar features can be seen. With the exception of crystal
C001 the intensity of \(\gamma\)-ray interactions is clearly seen to
be lower at segment boundaries. This is interpreted as coming from
the one-interaction per segment approximation used in the PSA and
supported by the results of performing PSA on calculated signals as
previously shown in figures \ref{fig:org8a5bc51} and \ref{fig:org87182c5}. Based on
the number of counts in the maximum bin the quality of the PSA
improves when moving from ADL bases to optimized AGATAGeFEM bases for
four out of six detectors. This is seen when comparing the scales
in figure \ref{fig:org92809e3} and figure \ref{fig:org602c4f7}. However, the "hot spots"
seen for crystal C001 remains pronounced for all bases. They are
located at the corners of the front face and for depths in the
crystal that is smaller than 4 mm (as determined using the ADL
bases). To investigate these events closer an average of all traces
belonging to the hot spot close to x\textasciitilde{}-30 mm, y\textasciitilde{}0 mm, z<4 mm was
produced together with an average of events that gave x and y
coordinates next to the hot spot (the two regions are marked in
figure \ref{fig:org92809e3}). On the events used for averaging the condition of
a net-charge in only one segment was also enforced. In figure
\ref{fig:orgecd679f} the resulting averages are shown together with the
pulse shapes from the ADL and optimized AGATAGeFEM bases coming from
positions in the bases corresponding to the hot spot. As can be seen
the average trace for events ending up at the hot spot does not reach
its full value. This is most likely related to an incorrect
determination of the start time for these events (in figure
\ref{fig:orgecd679f} the traces have been aligned for clarity). When trying
to fit the time-misaligned traces the PSA algorithm always finds the
same best position as the time alignment can be compensated by
choosing an extreme rise time in the pulse-shape data basis. When
comparing the traces from the ADL bases and the optimized AGATAGeFEM
bases an explanation to why the C001 hot spots are "less hot" for the
AGATAGeFEM bases is given. The start of the signal for the AGATAGeFEM
bases is different allowing for more positions to reproduce the
"false" rise time of the time-misaligned traces. It is however
difficult to state if one basis is more realistic than the other as
the "hot spot" is related to the preprocessing of the traces
performed before the PSA or to the time-pickup made in the
digitizers of AGATA. This underlines the importance of preprocessing
for successful PSA.

\section{Conclusions}
\label{sec:org3b533d1}
\label{org368ee3f} The C++ based software package AGATAGeFEM aiming
at modeling segmented High-Purity Germanium detectors has been
described. It allows the implementation of the detector geometry and
segmentation schemes to within machine precision and uses Finite
Element Methods to solve the Laplace and Poisson equations. The
resulting fields are calculated using the basis functions and
support points of the actual FEM grid, i.e. using function
evaluation rather than interpolation.

The charge-transport equations are solved using time adaptive
Runge-Kutta methods from the GNU Scientific Library. To the induced
charge signals linear and differential crosstalk is added together
with the transfer function of the electronics. Convolution is made
in the time domain. The model used in AGATAGeFEM for hole-charge
carrier velocity has proven to give good results for PSA
but still has room for improvements.

In this work AGATAGeFEM has been used to investigate the impact of
crosstalk and noise for the EGS PSA and for the SVD PSA. The result
suggests that crosstalk at the level of what is found in AGATA has a
small impact on the resolution of the PSA. Furthermore the
influence of an imperfectly known crystal geometry has been
investigated. It was found that a \(\chi^2\) figure-of-merit stating
on how good the experimental signals could be fitted using the
pulse-shape basis is not a good indicator for the precision of the
geometry. In extreme cases the measured position resolution using
in-beam methods can give indications. These results point to the
importance to have well defined crystal geometries when modeling
pulse shapes.

As a validation of the pulse-shapes calculated using AGATAGeFEM
pulse-shape data bases for PSA on AGATA data have been produced and
optimized. The resulting bases allow for analysing the data with
results that are as good as the other state-of-the-art pulse-shape
data bases, showing that the concepts and models used in AGATAGeFEM
are producing pulse shapes as realistic as the ADL presently used
within the AGATA collaboration. It has also been shown that "hot
spots" seen in the distribution of \(\gamma\)-ray interaction points
from the AGATA PSA can be linked to problems in the data treatment
prior to PSA, e.g. the time alignment of the traces.

\section*{Acknowledgement}
\label{sec:org620dbc0}
The author would like to thank the AGATA collaboration and the
GANIL technical staff. Gilbert Duchêne is thanked for providing the
in-beam data set used to extract the position resolution of the
pulse-shape analysis. R.M. Pérez-Vidal, A. Lopez-Martens,
C. Michelagnoli, E. Clément, J. Dudouet, and H. J. Li are thanked
for their contribution given via the work performed in the scope of
earlier publications. The excellent performance of the AGATA
detectors is assured by the AGATA Detector Working group. The AGATA
project is supported in France by the CNRS and the CEA. This work
has been supported by the OASIS project no. ANR-17-CE31-0026.

\appendix
\section{Parameters used for pulse-shape calculations}
\label{sec:orgbecd85b}

\label{org56d91ba}
\begin{table}[htbp]
\caption{\label{tab:org047b7d6}
Charge carrier mobility parameters used in this work}
\centering
\begin{tabular}{|c|c|}
Parameter & Value\\
\hline
\(\mu_{e}^{<100>}\) & 37200 \(\frac{cm^2}{Vs}\)\\
\(\beta_{e}^{<100>}\) & 0.805\\
\(E^{<100>}_{e0}\) & 510 \(\frac{V}{cm}\)\\
\(\mu_{en}^{<100>}\) & -167 \(\frac{cm^2}{Vs}\)\\
\(\mu_{e}^{<111>}\) & 32908 \(\frac{cm^2}{Vs}\)\\
\(\beta_{e}^{<111>}\) & 0.774\\
\(E^{<111>}_{e0}\) & 448 \(\frac{V}{cm}\)\\
\(\mu_{en}^{e<111>}\) & -133 \(\frac{cm^2}{Vs}\)\\
\(\mu_{h}^{<100>}\) & 62380 \(\frac{cm^2}{Vs}\)\\
\(\beta_{h}^{<100>}\) & 0.727\\
\(E^{<100>}_{h0}\) & 181 \(\frac{V}{cm}\)\\
\(\mu_{hn}^{<100>}\) & 0 \(\frac{cm^2}{Vs}\)\\
\(\mu_{h}^{<111>}\) & 62508 \(\frac{cm^2}{Vs}\)\\
\(\beta_{h}^{<111>}\) & 0.757\\
\(E^{<111>}_{h0}\) & 144 \(\frac{V}{cm}\)\\
\(\mu_{hn}^{<111>}\) & 0 \(\frac{cm^2}{Vs}\)\\
\end{tabular}
\end{table}

\begin{table}[htbp]
\caption{\label{tab:orge809380}
Space charge densities used for field calculations}
\centering
\begin{tabular}{|c|c|c|}
Crystal & Front [\(10^{10}/cm^{3}\)] & Back [\(10^{10}/cm^{3}\)]\\
\hline
A002 & 0.50 & 1.18\\
B010 & 0.54 & 1.55\\
C001 & 0.93 & 0.67\\
A007 & 0.42 & 1.59\\
B007 & 0.52 & 1.76\\
C007 & 0.60 & 1.39\\
\end{tabular}
\end{table}

\bibliographystyle{elsarticle-num}
\bibliography{../../../Mercurial/Ljungvall/referenser/referenser}

\begin{thebibliography}{10}
\expandafter\ifx\csname url\endcsname\relax
  \def\url#1{\texttt{#1}}\fi
\expandafter\ifx\csname urlprefix\endcsname\relax\def\urlprefix{URL }\fi
\expandafter\ifx\csname href\endcsname\relax
  \def\href#1#2{#2} \def\path#1{#1}\fi

\bibitem{Simpson1997}
J.~Simpson, \href{http://dx.doi.org/10.1007/s002180050290}{The euroball
  spectrometer}, Zeitschrift f{\"u}r Physik A Hadrons and Nuclei 358~(2) (1997)
  139--143.
\newblock \href {https://doi.org/10.1007/s002180050290}
  {\path{doi:10.1007/s002180050290}}.
\newline\urlprefix\url{http://dx.doi.org/10.1007/s002180050290}

\bibitem{Deleplanque1987}
M.~A. Deleplanque, R.~M.~e. Diamond, 1987 gammasphere proposal preprint
  lbnl-5202, Tech. rep., LBNL (1987).

\bibitem{Lee1999195}
I.~Lee,
  \href{http://www.sciencedirect.com/science/article/pii/S0168900298010936}{Gamma-ray
  tracking detectors}, Nuclear Instruments and Methods in Physics Research
  Section A: Accelerators, Spectrometers, Detectors and Associated Equipment
  422~(1-3) (1999) 195 -- 200.
\newblock \href
  {https://doi.org/http://dx.doi.org/10.1016/S0168-9002(98)01093-6}
  {\path{doi:http://dx.doi.org/10.1016/S0168-9002(98)01093-6}}.
\newline\urlprefix\url{http://www.sciencedirect.com/science/article/pii/S0168900298010936}

\bibitem{Akkoyun201226}
S.~Akkoyun, A.~Algora, B.~Alikhani, F.~Ameil, G.~de~Angelis, L.~Arnold,
  A.~Astier, A.~Ata\c{c}, Y.~Aubert, C.~Aufranc, A.~Austin, S.~Aydin,
  F.~Azaiez, S.~Badoer, D.~Balabanski, D.~Barrientos, G.~Baulieu, R.~Baumann,
  D.~Bazzacco, F.~Beck, T.~Beck, P.~Bednarczyk, M.~Bellato, M.~Bentley,
  G.~Benzoni, R.~Berthier, L.~Berti, R.~Beunard, G.~L. Bianco, B.~Birkenbach,
  P.~Bizzeti, A.~Bizzeti-Sona, F.~L. Blanc, J.~Blasco, N.~Blasi, D.~Bloor,
  C.~Boiano, M.~Borsato, D.~Bortolato, A.~Boston, H.~Boston, P.~Bourgault,
  P.~Boutachkov, A.~Bouty, A.~Bracco, S.~Brambilla, I.~Brawn, A.~Brondi,
  S.~Broussard, B.~Bruyneel, D.~Bucurescu, I.~Burrows, A.~B{\"u}rger,
  S.~Cabaret, B.~Cahan, E.~Calore, F.~Camera, A.~Capsoni, F.~Carri\'{o},
  G.~Casati, M.~Castoldi, B.~Cederwall, J.-L. Cercus, V.~Chambert, M.~E.
  Chambit, R.~Chapman, L.~Charles, J.~Chavas, E.~Cl\'{e}ment, P.~Cocconi,
  S.~Coelli, P.~Coleman-Smith, A.~Colombo, S.~Colosimo, C.~Commeaux,
  D.~Conventi, R.~Cooper, A.~Corsi, A.~Cortesi, L.~Costa, F.~Crespi,
  J.~Cresswell, D.~Cullen, D.~Curien, A.~Czermak, D.~Delbourg, R.~Depalo,
  T.~Descombes, P.~D\'{e}sesquelles, P.~Detistov, C.~Diarra, F.~Didierjean,
  M.~Dimmock, Q.~Doan, C.~Domingo-Pardo, M.~Doncel, F.~Dorangeville, N.~Dosme,
  Y.~Drouen, G.~Duch\^{e}ne, B.~Dulny, J.~Eberth, P.~Edelbruck, J.~Egea,
  T.~Engert, M.~Erduran, S.~Ert{\"u}rk, C.~Fanin, S.~Fantinel, E.~Farnea,
  T.~Faul, M.~Filliger, F.~Filmer, C.~Finck, G.~de~France, A.~Gadea, W.~Gast,
  A.~Geraci, J.~Gerl, R.~Gernh{\"a}user, A.~Giannatiempo, A.~Giaz, L.~Gibelin,
  A.~Givechev, N.~Goel, V.~Gonz\'{a}lez, A.~Gottardo, X.~Grave, J.~Grebosz,
  R.~Griffiths, A.~Grint, P.~Gros, L.~Guevara, M.~Gulmini, A.~G{\"o}rgen,
  H.~Ha, T.~Habermann, L.~Harkness, H.~Harroch, K.~Hauschild, C.~He,
  A.~Hern\'{a}ndez-Prieto, B.~Hervieu, H.~Hess, T.~H{\"u}y{\"u}k, E.~Ince,
  R.~Isocrate, G.~Jaworski, A.~Johnson, J.~Jolie, P.~Jones, B.~Jonson,
  P.~Joshi, D.~Judson, A.~Jungclaus, M.~Kaci, N.~Karkour, M.~Karolak,
  A.~Ka\k{s}ka\k{s}, M.~Kebbiri, R.~Kempley, A.~Khaplanov, S.~Klupp,
  M.~Kogimtzis, I.~Kojouharov, A.~Korichi, W.~Korten, T.~Kr{\"o}ll,
  R.~Kr{\"u}cken, N.~Kurz, B.~Ky, M.~Labiche, X.~Lafay, L.~Lavergne,
  I.~Lazarus, S.~Leboutelier, F.~Lefebvre, E.~Legay, L.~Legeard, F.~Lelli,
  S.~Lenzi, S.~Leoni, A.~Lermitage, D.~Lersch, J.~Leske, S.~Letts, S.~Lhenoret,
  R.~Lieder, D.~Linget, J.~Ljungvall, A.~Lopez-Martens, A.~Lotod\'{e},
  S.~Lunardi, A.~Maj, J.~van~der Marel, Y.~Mariette, N.~Marginean,
  R.~Marginean, G.~Maron, A.~Mather, W.~Meczy\'{n}ski, V.~Mend\'{e}z,
  P.~Medina, B.~Melon, R.~Menegazzo, D.~Mengoni, E.~Merchan, L.~Mihailescu,
  C.~Michelagnoli, J.~Mierzejewski, L.~Milechina, B.~Million, K.~Mitev,
  P.~Molini, D.~Montanari, S.~Moon, F.~Morbiducci, R.~Moro, P.~Morrall,
  O.~M{\"o}ller, A.~Nannini, D.~Napoli, L.~Nelson, M.~Nespolo, V.~Ngo,
  M.~Nicoletto, R.~Nicolini, Y.~L. Noa, P.~Nolan, M.~Norman, J.~Nyberg,
  A.~Obertelli, A.~Olariu, R.~Orlandi, D.~Oxley, C.~{\"O}zben, M.~Ozille,
  C.~Oziol, E.~Pachoud, M.~Palacz, J.~Palin, J.~Pancin, C.~Parisel, P.~Pariset,
  G.~Pascovici, R.~Peghin, L.~Pellegri, A.~Perego, S.~Perrier, M.~Petcu,
  P.~Petkov, C.~Petrache, E.~Pierre, N.~Pietralla, S.~Pietri, M.~Pignanelli,
  I.~Piqueras, Z.~Podolyak, P.~L. Pouhalec, J.~Pouthas, D.~Pugn\'{e}re,
  V.~Pucknell, A.~Pullia, B.~Quintana, R.~Raine, G.~Rainovski, L.~Ramina,
  G.~Rampazzo, G.~L. Rana, M.~Rebeschini, F.~Recchia, N.~Redon, M.~Reese,
  P.~Reiter, P.~Regan, S.~Riboldi, M.~Richer, M.~Rigato, S.~Rigby,
  G.~Ripamonti, A.~Robinson, J.~Robin, J.~Roccaz, J.-A. Ropert, B.~Ross\'{e},
  C.~R. Alvarez, D.~Rosso, B.~Rubio, D.~Rudolph, F.~Saillant, E.~\k'{S}ahin,
  F.~Salomon, M.-D. Salsac, J.~Salt, G.~Salvato, J.~Sampson, E.~Sanchis,
  C.~Santos, H.~Schaffner, M.~Schlarb, D.~Scraggs, D.~Seddon,
  M.~\k{S}enyi\^{g}it, M.-H. Sigward, G.~Simpson, J.~Simpson, M.~Slee,
  J.~Smith, P.~Sona, B.~Sowicki, P.~Spolaore, C.~Stahl, T.~Stanios,
  E.~Stefanova, O.~St\'{e}zowski, J.~Strachan, G.~Suliman, P.-A.
  S{\"o}derstr{\"o}m, J.~Tain, S.~Tanguy, S.~Tashenov, C.~Theisen,
  J.~Thornhill, F.~Tomasi, N.~Toniolo, R.~Touzery, B.~Travers, A.~Triossi,
  M.~Tripon, K.~Tun-Lano{\"e}, M.~Turcato, C.~Unsworth, C.~Ur,
  J.~Valiente-Dobon, V.~Vandone, E.~Vardaci, R.~Venturelli, F.~Veronese,
  C.~Veyssiere, E.~Viscione, R.~Wadsworth, P.~Walker, N.~Warr, C.~Weber,
  D.~Weisshaar, D.~Wells, O.~Wieland, A.~Wiens, G.~Wittwer, H.~Wollersheim,
  F.~Zocca, N.~Zamfir, M.~Ziebli\'{n}ski, A.~Zucchiatti,
  \href{http://www.sciencedirect.com/science/article/pii/S0168900211021516}{Agata-advanced
  \{GAmma\} tracking array}, Nuclear Instruments and Methods in Physics
  Research Section A: Accelerators, Spectrometers, Detectors and Associated
  Equipment 668 (2012) 26 -- 58.
\newblock \href {https://doi.org/http://dx.doi.org/10.1016/j.nima.2011.11.081}
  {\path{doi:http://dx.doi.org/10.1016/j.nima.2011.11.081}}.
\newline\urlprefix\url{http://www.sciencedirect.com/science/article/pii/S0168900211021516}

\bibitem{Lee20031095}
I.~Y. Lee, M.~A. Deleplanque, K.~Vetter,
  \href{http://stacks.iop.org/0034-4885/66/i=7/a=201}{Developments in large
  gamma-ray detector arrays}, Reports on Progress in Physics 66~(7) (2003)
  1095.
\newline\urlprefix\url{http://stacks.iop.org/0034-4885/66/i=7/a=201}

\bibitem{EBERTH2008283}
J.~Eberth, J.~Simpson,
  \href{http://www.sciencedirect.com/science/article/pii/S0146641007000828}{From
  ge(li) detectors to gamma-ray tracking arrays--50 years of gamma spectroscopy
  with germanium detectors}, Progress in Particle and Nuclear Physics 60~(2)
  (2008) 283 -- 337.
\newblock \href {https://doi.org/https://doi.org/10.1016/j.ppnp.2007.09.001}
  {\path{doi:https://doi.org/10.1016/j.ppnp.2007.09.001}}.
\newline\urlprefix\url{http://www.sciencedirect.com/science/article/pii/S0146641007000828}

\bibitem{LJUNGVALL2020163297}
J.~Ljungvall, R.~P{\'e}rez-Vidal, A.~Lopez-Martens, C.~Michelagnoli,
  E.~Cl{\'e}ment, J.~Dudouet, A.~Gadea, H.~Hess, A.~Korichi, M.~Labiche,
  N.~Lalovi{\'c}, H.~Li, F.~Recchia,
  \href{http://www.sciencedirect.com/science/article/pii/S0168900219315475}{Performance
  of the advanced gamma tracking array at ganil}, Nuclear Instruments and
  Methods in Physics Research Section A: Accelerators, Spectrometers, Detectors
  and Associated Equipment 955 (2020) 163297.
\newblock \href {https://doi.org/https://doi.org/10.1016/j.nima.2019.163297}
  {\path{doi:https://doi.org/10.1016/j.nima.2019.163297}}.
\newline\urlprefix\url{http://www.sciencedirect.com/science/article/pii/S0168900219315475}

\bibitem{Schlarb20111}
M.~Schlarb, R.~Gernh{\"a}user, S.~Klupp, R.~Kr{\"u}cken,
  \href{http://dx.doi.org/10.1140/epja/i2011-11131-3}{Pulse shape analysis for
  $\gamma$-ray tracking (part ii): Fully informed particle swarm algorithm
  applied to agata}, The European Physical Journal A 47~(10) (2011) 1--9.
\newblock \href {https://doi.org/10.1140/epja/i2011-11131-3}
  {\path{doi:10.1140/epja/i2011-11131-3}}.
\newline\urlprefix\url{http://dx.doi.org/10.1140/epja/i2011-11131-3}

\bibitem{Schlarb2011}
M.~Schlarb, R.~Gernh{\"a}user, S.~Klupp, R.~Kr{\"u}cken,
  \href{http://dx.doi.org/10.1140/epja/i2011-11132-2}{Pulse shape analysis for
  $\gamma$-ray tracking (part i): Pulse shape simulation with jass}, The
  European Physical Journal A 47~(10) (2011) 132.
\newblock \href {https://doi.org/10.1140/epja/i2011-11132-2}
  {\path{doi:10.1140/epja/i2011-11132-2}}.
\newline\urlprefix\url{http://dx.doi.org/10.1140/epja/i2011-11132-2}

\bibitem{Matue2014}
I.~Mateu, P.~Medina, J.~Roques, E.~Jourdain, Simulation of the charge
  collection and signal response of a hpge double sided strip detector using
  mgs, Nuclear Instruments and Methods in Physics Research A 735 (2014)
  574--583.
\newblock \href {https://doi.org/10.1016/j.nima.2013.09.069}
  {\path{doi:10.1016/j.nima.2013.09.069}}.

\bibitem{Bruyneel20161}
B.~Bruyneel, B.~Birkenbach, P.~Reiter,
  \href{http://dx.doi.org/10.1140/epja/i2016-16070-9}{Pulse shape analysis and
  position determination in segmented hpge detectors: The agata detector
  library}, The European Physical Journal A 52~(3) (2016) 1--11.
\newblock \href {https://doi.org/10.1140/epja/i2016-16070-9}
  {\path{doi:10.1140/epja/i2016-16070-9}}.
\newline\urlprefix\url{http://dx.doi.org/10.1140/epja/i2016-16070-9}

\bibitem{Bruyneel2016}
B.~Bruyneel, B.~Birkenbach, P.~Reiter,
  \href{http://dx.doi.org/10.1140/epja/i2016-16070-9}{Pulse shape analysis and
  position determination in segmented hpge detectors: The agata detector
  library}, The European Physical Journal A 52~(3) (2016) 70.
\newblock \href {https://doi.org/10.1140/epja/i2016-16070-9}
  {\path{doi:10.1140/epja/i2016-16070-9}}.
\newline\urlprefix\url{http://dx.doi.org/10.1140/epja/i2016-16070-9}

\bibitem{BIRKENBACH2011176}
B.~Birkenbach, B.~Bruyneel, G.~Pascovici, J.~Eberth, H.~Hess, D.~Lersch,
  P.~Reiter, A.~Wiens,
  \href{http://www.sciencedirect.com/science/article/pii/S0168900211005912}{Determination
  of space charge distributions in highly segmented large volume hpge detectors
  from capacitance–voltage measurements}, Nuclear Instruments and Methods in
  Physics Research Section A: Accelerators, Spectrometers, Detectors and
  Associated Equipment 640~(1) (2011) 176 -- 184.
\newblock \href {https://doi.org/https://doi.org/10.1016/j.nima.2011.02.109}
  {\path{doi:https://doi.org/10.1016/j.nima.2011.02.109}}.
\newline\urlprefix\url{http://www.sciencedirect.com/science/article/pii/S0168900211005912}

\bibitem{BRUYNEEL201192}
B.~Bruyneel, B.~Birkenbach, P.~Reiter,
  \href{http://www.sciencedirect.com/science/article/pii/S0168900211006085}{Space
  charge reconstruction in highly segmented hpge detectors through
  capacitance-voltage measurements}, Nuclear Instruments and Methods in Physics
  Research Section A: Accelerators, Spectrometers, Detectors and Associated
  Equipment 641~(1) (2011) 92 -- 100.
\newblock \href {https://doi.org/https://doi.org/10.1016/j.nima.2011.02.110}
  {\path{doi:https://doi.org/10.1016/j.nima.2011.02.110}}.
\newline\urlprefix\url{http://www.sciencedirect.com/science/article/pii/S0168900211006085}

\bibitem{Bruyneel2009196}
B.~Bruyneel, P.~Reiter, A.~Wiens, J.~Eberth, H.~Hess, G.~Pascovici, N.~Warr,
  D.~Weisshaar,
  \href{http://www.sciencedirect.com/science/article/pii/S0168900208015921}{Crosstalk
  properties of 36-fold segmented symmetric hexagonal \{HPGe\} detectors},
  Nuclear Instruments and Methods in Physics Research Section A: Accelerators,
  Spectrometers, Detectors and Associated Equipment 599~(2-3) (2009) 196 --
  208.
\newblock \href {https://doi.org/http://dx.doi.org/10.1016/j.nima.2008.11.011}
  {\path{doi:http://dx.doi.org/10.1016/j.nima.2008.11.011}}.
\newline\urlprefix\url{http://www.sciencedirect.com/science/article/pii/S0168900208015921}

\bibitem{BRUYNEEL200999}
B.~Bruyneel, P.~Reiter, A.~Wiens, J.~Eberth, H.~Hess, G.~Pascovici, N.~Warr,
  S.~Aydin, D.~Bazzacco, F.~Recchia,
  \href{http://www.sciencedirect.com/science/article/pii/S0168900209012455}{Crosstalk
  corrections for improved energy resolution with highly segmented
  hpge-detectors}, Nuclear Instruments and Methods in Physics Research Section
  A: Accelerators, Spectrometers, Detectors and Associated Equipment 608~(1)
  (2009) 99 -- 106.
\newblock \href {https://doi.org/https://doi.org/10.1016/j.nima.2009.06.037}
  {\path{doi:https://doi.org/10.1016/j.nima.2009.06.037}}.
\newline\urlprefix\url{http://www.sciencedirect.com/science/article/pii/S0168900209012455}

\bibitem{schlarb2008simulation}
M.~Schlarb, R.~Gernh{\"a}user, R.~Kr{\"u}cken, Simulation and real-time
  analysis of pulse shapes from hpge detectors, Tech. rep. (2008).

\bibitem{Bruyneel2006764}
B.~Bruyneel, P.~Reiter, G.~Pascovici,
  \href{http://www.sciencedirect.com/science/article/pii/S0168900206015166}{Characterization
  of large volume \{HPGe\} detectors. part i: Electron and hole mobility
  parameterization}, Nuclear Instruments and Methods in Physics Research
  Section A: Accelerators, Spectrometers, Detectors and Associated Equipment
  569~(3) (2006) 764 -- 773.
\newblock \href {https://doi.org/http://dx.doi.org/10.1016/j.nima.2006.08.130}
  {\path{doi:http://dx.doi.org/10.1016/j.nima.2006.08.130}}.
\newline\urlprefix\url{http://www.sciencedirect.com/science/article/pii/S0168900206015166}

\bibitem{Paschalis201344}
S.~Paschalis, I.~Lee, A.~Macchiavelli, C.~Campbell, M.~Cromaz, S.~Gros,
  J.~Pavan, J.~Qian, R.~Clark, H.~Crawford, D.~Doering, P.~Fallon,
  C.~Lionberger, T.~Loew, M.~Petri, T.~Stezelberger, S.~Zimmermann, D.~Radford,
  K.~Lagergren, D.~Weisshaar, R.~Winkler, T.~Glasmacher, J.~Anderson,
  C.~Beausang,
  \href{http://www.sciencedirect.com/science/article/pii/S0168900213000508}{The
  performance of the gamma-ray energy tracking in-beam nuclear array gretina},
  Nuclear Instruments and Methods in Physics Research Section A: Accelerators,
  Spectrometers, Detectors and Associated Equipment 709 (2013) 44 -- 55.
\newblock \href {https://doi.org/https://doi.org/10.1016/j.nima.2013.01.009}
  {\path{doi:https://doi.org/10.1016/j.nima.2013.01.009}}.
\newline\urlprefix\url{http://www.sciencedirect.com/science/article/pii/S0168900213000508}

\bibitem{PRASHER201750}
V.~Prasher, M.~Cromaz, E.~Merchan, P.~Chowdhury, H.~Crawford, C.~Lister,
  C.~Campbell, I.~Lee, A.~Macchiavelli, D.~Radford, A.~Wiens,
  \href{http://www.sciencedirect.com/science/article/pii/S0168900216311925}{Sensitivity
  of gretina position resolution to hole mobility}, Nuclear Instruments and
  Methods in Physics Research Section A: Accelerators, Spectrometers, Detectors
  and Associated Equipment 846 (2017) 50 -- 55.
\newblock \href {https://doi.org/https://doi.org/10.1016/j.nima.2016.11.038}
  {\path{doi:https://doi.org/10.1016/j.nima.2016.11.038}}.
\newline\urlprefix\url{http://www.sciencedirect.com/science/article/pii/S0168900216311925}

\bibitem{Ramo1939}
S.~Ramo, Proc. IRE 27 (1939) 584.

\bibitem{Shockley1938}
W.~Shockley, J. Appl. Phys. 9 (1938) 635.

\bibitem{knoll2000}
G.~Knoll, Radiation Detection and Measurement, Wiley, 2000.

\bibitem{Nathan1963}
M.~I. Nathan,
  \href{http://link.aps.org/doi/10.1103/PhysRev.130.2201}{Anisotropy of the
  conductivity of $n$-type germanium at high electric fields}, Phys. Rev. 130
  (1963) 2201--2204.
\newblock \href {https://doi.org/10.1103/PhysRev.130.2201}
  {\path{doi:10.1103/PhysRev.130.2201}}.
\newline\urlprefix\url{http://link.aps.org/doi/10.1103/PhysRev.130.2201}

\bibitem{Mihailescu2000350}
L.~Mihailescu, W.~Gast, R.~Lieder, H.~Brands, H.~J{\"a}ger,
  \href{http://www.sciencedirect.com/science/article/pii/S0168900299012863}{The
  influence of anisotropic electron drift velocity on the signal shapes of
  closed-end \{HPGe\} detectors}, Nuclear Instruments and Methods in Physics
  Research Section A: Accelerators, Spectrometers, Detectors and Associated
  Equipment 447~(3) (2000) 350 -- 360.
\newblock \href
  {https://doi.org/http://dx.doi.org/10.1016/S0168-9002(99)01286-3}
  {\path{doi:http://dx.doi.org/10.1016/S0168-9002(99)01286-3}}.
\newline\urlprefix\url{http://www.sciencedirect.com/science/article/pii/S0168900299012863}

\bibitem{LjungvallThesis}
.~Ljungvall, Joa, U.~universitet, F.~sektionen, I.~f{\"o}r k{\"a}rn-och
  partikelfysik, T.~naturvetenskapliga vetenskapsomr{\aa}det, Characterisation
  of the neutron wall and of neutron interactions in germanium-detector systems
  (2005).

\bibitem{kittel1996}
C.~Kittel, \href{https://books.google.fr/books?id=1X8pAQAAMAAJ}{Introduction to
  Solid State Physics}, Wiley, 1996.
\newline\urlprefix\url{https://books.google.fr/books?id=1X8pAQAAMAAJ}

\bibitem{Mei2016}
H.~Mei, D.-M. Mei, G.-J. Wang, G.~Yang,
  \href{https://doi.org/10.1088%2F1748-0221%2F11%2F12%2Fp12021}{The impact of
  neutral impurity concentration on charge drift mobility in p-type germanium},
  Journal of Instrumentation 11~(12) (2016) P12021--P12021.
\newblock \href {https://doi.org/10.1088/1748-0221/11/12/p12021}
  {\path{doi:10.1088/1748-0221/11/12/p12021}}.
\newline\urlprefix\url{https://doi.org/10.1088%2F1748-0221%2F11%2F12%2Fp12021}

\bibitem{Lewandowski2019}
L.~Lewandowski, P.~Reiter, B.~Birkenbach, B.~Bruyneel, E.~Clement, J.~Eberth,
  H.~Hess, C.~Michelagnoli, H.~Li, R.~M. Perez-Vidal, M.~Zielinska,
  \href{https://doi.org/10.1140/epja/i2019-12752-0}{Pulse-shape analysis and
  position resolution in highly segmented hpge agata detectors}, The European
  Physical Journal A 55~(5) (2019) 81.
\newblock \href {https://doi.org/10.1140/epja/i2019-12752-0}
  {\path{doi:10.1140/epja/i2019-12752-0}}.
\newline\urlprefix\url{https://doi.org/10.1140/epja/i2019-12752-0}

\bibitem{Medina}
\href{http://www.iphc.cnrs.fr/-MGS-.html}{[link]}.
\newline\urlprefix\url{http://www.iphc.cnrs.fr/-MGS-.html}

\bibitem{BrennerFEM}
S.~C. Brenner, L.~R. Scott, The Mathematical Theory of Finite ElementMethods,
  2nd Edition, Vol.~15 of Texts in Applied Mathematics, Springer, 2002.

\bibitem{gsl2010}
G.~P. Contributors, \href{http://www.gnu.org/software/gsl/}{{GSL} - {GNU}
  scientific library - {GNU} project - free software foundation {(FSF)}},
  http://www.gnu.org/software/gsl/ (2010) [cited 2010-06-20 22:49:12].
\newline\urlprefix\url{http://www.gnu.org/software/gsl/}

\bibitem{BangerthHartmannKanschat2007}
W.~Bangerth, R.~Hartmann, G.~Kanschat, {deal.II} -- a general purpose object
  oriented finite element library, ACM Trans. Math. Softw. 33~(4) (2007)
  24/1--24/27.

\bibitem{dealII90}
G.~Alzetta, D.~Arndt, W.~Bangerth, V.~Boddu, B.~Brands, D.~Davydov,
  R.~Gassmoeller, T.~Heister, L.~Heltai, K.~Kormann, M.~Kronbichler, M.~Maier,
  J.-P. Pelteret, B.~Turcksin, D.~Wells, The \texttt{deal.II} library, version
  9.0, Journal of Numerical Mathematics 26~(4) (2018) 173--183.
\newblock \href {https://doi.org/10.1515/jnma-2018-0054}
  {\path{doi:10.1515/jnma-2018-0054}}.

\bibitem{Kirk2006237}
B.~S. Kirk, J.~W. Peterson, R.~H. Stogner, G.~F. Carey, {\texttt{libMesh}: A
  C++ Library for Parallel Adaptive Mesh Refinement/Coarsening Simulations},
  Engineering with Computers 22~(3--4) (2006) 237--254,
  \url{http://dx.doi.org/10.1007/s00366-006-0049-3}.

\bibitem{Hatlo2005}
M.~Hatlo, F.~James, P.~Mato, L.~Moneta, M.~Winkler, A.~Zsenei, Developments of
  mathematical software libraries for the lhc experiments, IEEE Transactions on
  Nuclear Science 52~(6) (2005) 2818--2822.
\newblock \href {https://doi.org/10.1109/TNS.2005.860152}
  {\path{doi:10.1109/TNS.2005.860152}}.

\bibitem{ANTCHEVA20092499}
I.~Antcheva, M.~Ballintijn, B.~Bellenot, M.~Biskup, R.~Brun, N.~Buncic,
  P.~Canal, D.~Casadei, O.~Couet, V.~Fine, L.~Franco, G.~Ganis, A.~Gheata,
  D.~G. Maline, M.~Goto, J.~Iwaszkiewicz, A.~Kreshuk, D.~M. Segura, R.~Maunder,
  L.~Moneta, A.~Naumann, E.~Offermann, V.~Onuchin, S.~Panacek, F.~Rademakers,
  P.~Russo, M.~Tadel,
  \href{http://www.sciencedirect.com/science/article/pii/S0010465509002550}{Root
  — a c++ framework for petabyte data storage, statistical analysis and
  visualization}, Computer Physics Communications 180~(12) (2009) 2499 -- 2512,
  40 YEARS OF CPC: A celebratory issue focused on quality software for high
  performance, grid and novel computing architectures.
\newblock \href {https://doi.org/https://doi.org/10.1016/j.cpc.2009.08.005}
  {\path{doi:https://doi.org/10.1016/j.cpc.2009.08.005}}.
\newline\urlprefix\url{http://www.sciencedirect.com/science/article/pii/S0010465509002550}

\bibitem{Farnea2010331}
E.~Farnea, F.~Recchia, D.~Bazzacco, T.~Kr{\"o}ll, Z.~Podoly\'{a}k, B.~Quintana,
  A.~Gadea,
  \href{http://www.sciencedirect.com/science/article/pii/S0168900210008922}{Conceptual
  design and monte carlo simulations of the \{AGATA\} array}, Nuclear
  Instruments and Methods in Physics Research Section A: Accelerators,
  Spectrometers, Detectors and Associated Equipment 621~(1-3) (2010) 331 --
  343.
\newblock \href {https://doi.org/http://dx.doi.org/10.1016/j.nima.2010.04.043}
  {\path{doi:http://dx.doi.org/10.1016/j.nima.2010.04.043}}.
\newline\urlprefix\url{http://www.sciencedirect.com/science/article/pii/S0168900210008922}

\bibitem{Agostinelli2003250}
S.~Agostinelli, J.~Allison, K.~Amako, J.~Apostolakis, H.~Araujo, P.~Arce,
  M.~Asai, D.~Axen, S.~Banerjee, G.~Barrand, F.~Behner, L.~Bellagamba,
  J.~Boudreau, L.~Broglia, A.~Brunengo, H.~Burkhardt, S.~Chauvie, J.~Chuma,
  R.~Chytracek, G.~Cooperman, G.~Cosmo, P.~Degtyarenko, A.~Dell'Acqua,
  G.~Depaola, D.~Dietrich, R.~Enami, A.~Feliciello, C.~Ferguson, H.~Fesefeldt,
  G.~Folger, F.~Foppiano, A.~Forti, S.~Garelli, S.~Giani, R.~Giannitrapani,
  D.~Gibin, J.~G. Cadenas, I.~González, G.~G. Abril, G.~Greeniaus, W.~Greiner,
  V.~Grichine, A.~Grossheim, S.~Guatelli, P.~Gumplinger, R.~Hamatsu,
  K.~Hashimoto, H.~Hasui, A.~Heikkinen, A.~Howard, V.~Ivanchenko, A.~Johnson,
  F.~Jones, J.~Kallenbach, N.~Kanaya, M.~Kawabata, Y.~Kawabata, M.~Kawaguti,
  S.~Kelner, P.~Kent, A.~Kimura, T.~Kodama, R.~Kokoulin, M.~Kossov,
  H.~Kurashige, E.~Lamanna, T.~Lampén, V.~Lara, V.~Lefebure, F.~Lei,
  M.~Liendl, W.~Lockman, F.~Longo, S.~Magni, M.~Maire, E.~Medernach,
  K.~Minamimoto, P.~M. de~Freitas, Y.~Morita, K.~Murakami, M.~Nagamatu,
  R.~Nartallo, P.~Nieminen, T.~Nishimura, K.~Ohtsubo, M.~Okamura, S.~O'Neale,
  Y.~Oohata, K.~Paech, J.~Perl, A.~Pfeiffer, M.~Pia, F.~Ranjard, A.~Rybin,
  S.~Sadilov, E.~D. Salvo, G.~Santin, T.~Sasaki, N.~Savvas, Y.~Sawada,
  S.~Scherer, S.~Sei, V.~Sirotenko, D.~Smith, N.~Starkov, H.~Stoecker,
  J.~Sulkimo, M.~Takahata, S.~Tanaka, E.~Tcherniaev, E.~S. Tehrani,
  M.~Tropeano, P.~Truscott, H.~Uno, L.~Urban, P.~Urban, M.~Verderi, A.~Walkden,
  W.~Wander, H.~Weber, J.~Wellisch, T.~Wenaus, D.~Williams, D.~Wright,
  T.~Yamada, H.~Yoshida, D.~Zschiesche,
  \href{http://www.sciencedirect.com/science/article/pii/S0168900203013688}{Geant4—a
  simulation toolkit}, Nuclear Instruments and Methods in Physics Research
  Section A: Accelerators, Spectrometers, Detectors and Associated Equipment
  506~(3) (2003) 250 -- 303.
\newblock \href {https://doi.org/https://doi.org/10.1016/S0168-9002(03)01368-8}
  {\path{doi:https://doi.org/10.1016/S0168-9002(03)01368-8}}.
\newline\urlprefix\url{http://www.sciencedirect.com/science/article/pii/S0168900203013688}

\bibitem{Aydin2007}
S.~Aydin, F.~Recchia, D.~Bazzacco, E.~Farnea, C.~Ur,
  \href{http://www.lnl.infn.it/~annrep/read_ar/2007/contributions/pdfs/195_FA_103_FAA098.pdf}{Effective
  size of segmentation lines of an agata crystal}, Laboratori Nazionali di
  Legnaro, Annual Report (2007) 195--196.
\newline\urlprefix\url{http://www.lnl.infn.it/~annrep/read_ar/2007/contributions/pdfs/195_FA_103_FAA098.pdf}

\bibitem{Geuzaine09}
C.~Geuzaine, J.-F. Remacle, Gmsh: {A 3-D} finite element mesh generator with
  built-in pre- and post-processing facilities, International Journal for
  Numerical Methods in Engineering 79 (2009) 1309 -- 1331.

\bibitem{Kelly1983}
D.~W. Kelly, J.~P. De~S. R.~Gago, O.~C. Zienkiewicz, I.~Babuska,
  \href{https://onlinelibrary.wiley.com/doi/abs/10.1002/nme.1620191103}{A
  posteriori error analysis and adaptive processes in the finite element
  method: Part i—error analysis}, International Journal for Numerical Methods
  in Engineering 19~(11) (1983) 1593--1619.
\newblock \href
  {http://arxiv.org/abs/https://onlinelibrary.wiley.com/doi/pdf/10.1002/nme.1620191103}
  {\path{arXiv:https://onlinelibrary.wiley.com/doi/pdf/10.1002/nme.1620191103}},
  \href {https://doi.org/10.1002/nme.1620191103}
  {\path{doi:10.1002/nme.1620191103}}.
\newline\urlprefix\url{https://onlinelibrary.wiley.com/doi/abs/10.1002/nme.1620191103}

\bibitem{DORMAND198019}
J.~Dormand, P.~Prince,
  \href{http://www.sciencedirect.com/science/article/pii/0771050X80900133}{A
  family of embedded runge-kutta formulae}, Journal of Computational and
  Applied Mathematics 6~(1) (1980) 19 -- 26.
\newblock \href {https://doi.org/https://doi.org/10.1016/0771-050X(80)90013-3}
  {\path{doi:https://doi.org/10.1016/0771-050X(80)90013-3}}.
\newline\urlprefix\url{http://www.sciencedirect.com/science/article/pii/0771050X80900133}

\bibitem{Wiens2010223}
A.~Wiens, H.~Hess, B.~Birkenbach, B.~Bruyneel, J.~Eberth, D.~Lersch,
  G.~Pascovici, P.~Reiter, H.-G. Thomas,
  \href{http://www.sciencedirect.com/science/article/pii/S0168900210003384}{The
  \{AGATA\} triple cluster detector}, Nuclear Instruments and Methods in
  Physics Research Section A: Accelerators, Spectrometers, Detectors and
  Associated Equipment 618~(1-3) (2010) 223 -- 233.
\newblock \href {https://doi.org/http://dx.doi.org/10.1016/j.nima.2010.02.102}
  {\path{doi:http://dx.doi.org/10.1016/j.nima.2010.02.102}}.
\newline\urlprefix\url{http://www.sciencedirect.com/science/article/pii/S0168900210003384}

\bibitem{Soderstrom201196}
P.-A. S{\"o}derstr{\"o}m, F.~Recchia, J.~Nyberg, A.~Al-Adili, A.~Ata\c{c},
  S.~Aydin, D.~Bazzacco, P.~Bednarczyk, B.~Birkenbach, D.~Bortolato, A.~Boston,
  H.~Boston, B.~Bruyneel, D.~Bucurescu, E.~Calore, S.~Colosimo, F.~Crespi,
  N.~Dosme, J.~Eberth, E.~Farnea, F.~Filmer, A.~Gadea, A.~Gottardo, X.~Grave,
  J.~Grebosz, R.~Griffiths, M.~Gulmini, T.~Habermann, H.~Hess, G.~Jaworski,
  P.~Jones, P.~Joshi, D.~Judson, R.~Kempley, A.~Khaplanov, E.~Legay, D.~Lersch,
  J.~Ljungvall, A.~Lopez-Martens, W.~Meczynski, D.~Mengoni, C.~Michelagnoli,
  P.~Molini, D.~Napoli, R.~Orlandi, G.~Pascovici, A.~Pullia, P.~Reiter,
  E.~Sahin, J.~Smith, J.~Strachan, D.~Tonev, C.~Unsworth, C.~Ur,
  J.~Valiente-Dob\'{o}n, C.~Veyssiere, A.~Wiens,
  \href{http://www.sciencedirect.com/science/article/pii/S016890021100489X}{Interaction
  position resolution simulations and in-beam measurements of the \{AGATA\}
  \{HPGe\} detectors}, Nuclear Instruments and Methods in Physics Research
  Section A: Accelerators, Spectrometers, Detectors and Associated Equipment
  638~(1) (2011) 96 -- 109.
\newblock \href {https://doi.org/http://dx.doi.org/10.1016/j.nima.2011.02.089}
  {\path{doi:http://dx.doi.org/10.1016/j.nima.2011.02.089}}.
\newline\urlprefix\url{http://www.sciencedirect.com/science/article/pii/S016890021100489X}

\bibitem{Desesquelles2009}
P.~D{\'e}sesquelles, T.~M.~H. Ha, K.~Hauschild, A.~Korichi, F.~Le~Blanc,
  A.~Lopez-Martens, A.~Olariu, C.~M. Petrache,
  \href{http://dx.doi.org/10.1140/epja/i2008-10749-4}{Matrix formalism and
  singular-value decomposition for the location of gamma interactions in
  segmented hpge detectors}, The European Physical Journal A 40~(2) (2009)
  237--248.
\newblock \href {https://doi.org/10.1140/epja/i2008-10749-4}
  {\path{doi:10.1140/epja/i2008-10749-4}}.
\newline\urlprefix\url{http://dx.doi.org/10.1140/epja/i2008-10749-4}

\bibitem{Dimmock2009}
M.~R. Dimmock, A.~J. Boston, J.~R. Cresswell, I.~Lazarus, P.~Medina, P.~Nolan,
  C.~Parisel, C.~Santos, J.~Simpson, C.~Unsworth, Validation of pulse shape
  simulations for an agata prototype detector, IEEE Transactions on Nuclear
  Science 56~(4) (2009) 2415--2425.
\newblock \href {https://doi.org/10.1109/TNS.2009.2021842}
  {\path{doi:10.1109/TNS.2009.2021842}}.

\bibitem{Dimmock20091}
M.~R. {Dimmock}, A.~J. {Boston}, H.~C. {Boston}, J.~R. {Cresswell},
  L.~{Nelson}, P.~J. {Nolan}, C.~{Unsworth}, I.~H. {Lazarus}, J.~{Simpson},
  Characterisation results from an agata prototype detector, IEEE Transactions
  on Nuclear Science 56~(3) (2009) 1593--1599.
\newblock \href {https://doi.org/10.1109/TNS.2009.2019103}
  {\path{doi:10.1109/TNS.2009.2019103}}.

\bibitem{ventrurelli2004}
D.~B. R.~Venturelli, Adaptive grid search as pulse shape analysis algorithm for
  $\gamma$-tracking and results, Tech. rep., LNL (2004).

\bibitem{Li2018}
H.~J. Li, J.~Ljungvall, C.~Michelagnoli, E.~Cl{\'e}ment, J.~Dudouet,
  P.~D{\'e}sesquelles, A.~Lopez-Martens, G.~de~France,
  \href{https://doi.org/10.1140/epja/i2018-12636-9}{Experimental determination
  of reference pulses for highly segmented hpge detectors and application to
  pulse shape analysis used in $\gamma$-ray tracking arrays}, The European
  Physical Journal A 54~(11) (2018) 198.
\newblock \href {https://doi.org/10.1140/epja/i2018-12636-9}
  {\path{doi:10.1140/epja/i2018-12636-9}}.
\newline\urlprefix\url{https://doi.org/10.1140/epja/i2018-12636-9}

\bibitem{RecchiPHD}
F.~Recchia, Ph.D. thesis, Universit{\`a} degli Studi di Padova (2008).

\end{thebibliography}
\end{document}